\documentclass[aps,twocolumn,superscriptaddress]{revtex4}
\usepackage[dvips]{graphicx}
\usepackage[]{caption}

\begin{document}
\bibliographystyle{apsrev}

\title{Exact Criterion for Determining Clustering vs.\ Reentrant Melting 
Behavior for Bounded Interaction Potentials}
	   
\author{C. N. Likos}
\email[e-mail address: ] {likos@thphy.uni-duesseldorf.de}
\affiliation{Institut f{\"u}r Theoretische Physik II,
Heinrich-Heine-Universit\"at D{\"u}sseldorf,
Universit\"atsstra{\ss}e 1, D-40225 D\"usseldorf, Germany}
\author{A. Lang}
\affiliation{Institut f{\"u}r Theoretische Physik II,
Heinrich-Heine-Universit\"at D{\"u}sseldorf,
Universit\"atsstra{\ss}e 1, D-40225 D\"usseldorf, Germany}
\affiliation{Institut f\"ur Theoretische Physik and CMS,
Technische Universit\"at Wien, Wiedner Hauptstra{\ss}e 8-10, A-1040
Wien, Austria}
\author{M. Watzlawek}
\altaffiliation{Current address: Bayer AG, Central Research Division,
D-51368 Leverkusen, Germany}
\affiliation{Institut f{\"u}r Theoretische Physik II,
Heinrich-Heine-Universit\"at D{\"u}sseldorf,
Universit\"atsstra{\ss}e 1, D-40225 D\"usseldorf, Germany}
\author{H. L{\"o}wen}
\affiliation{Institut f{\"u}r Theoretische Physik II,
Heinrich-Heine-Universit\"at D{\"u}sseldorf,
Universit\"atsstra{\ss}e 1, D-40225 D\"usseldorf, Germany}
\date{\today, submitted to Physical Review E}

\begin{abstract}
We examine in full generality the phase behavior of systems whose
constituent particles interact by means of potentials which do not
diverge at the origin, are free of attractive parts and decay fast enough
to zero as the interparticle separation $r$ goes to infinity. By employing
a mean field-density functional theory which is shown to become exact
at high temperatures and/or densities,
we establish a criterion 
which determines whether a given system will freeze at all temperatures
or it will display reentrant melting and an upper freezing temperature.
\end{abstract}

\maketitle

\section{Introduction}

The phase behavior of systems whose constituent particles interact by
means of pair potentials diverging at the origin is a problem that
has been extensively studied in the last few decades. The whole range of
inverse power-law
pair potentials have been examined, ranging from hard spheres (HS)
to the one-component plasma (OCP) and it has been established that 
excluded volume effects are mainly responsible for bringing about the
freezing transition. The 
crystal structure in which a liquid freezes is subsequently determined 
by the steepness of the repulsion, with hard repulsions favoring a
face centered cubic (fcc) lattice
and soft ones a body centered cubic
(bcc) lattice \cite{mcconnell:etal:prl:93}.
Power-law diverging potentials result into freezing at arbitrarily 
high temperatures. However, the divergence of the potential alone
is not enough to cause such a phenomenon, as demonstrated
recently by Watzlawek {\it et al.} \cite{watzlawek:etal:prl:99} who
employed a logarithmically divergent pair potential, suitable to
describe effective interactions between star polymers in good solvents.
It was shown that the strength (prefactor) of the logarithmic potential,
determined by the number of arms $f$ of the stars,
is crucial in determining whether the system freezes or remains fluid at
all densities. As a consequence, the phase diagram of star polymers
was predicted to display reentrant melting and a critical freezing value 
$f_c = 34$
of arms, such that for $f < f_c$ the system remains always 
fluid \cite{watzlawek:etal:prl:99}.

Another interesting class of interactions are those which do not 
diverge at the origin, i.e., they are {\it bounded}. Such potentials
arise naturally as {\it effective interactions} between the centers
of mass of soft, flexible macromolecules such as polymer 
chains \cite{ard:peter:00},
dendrimers \cite{likos:ballauff}, polyelectrolytes etc. Indeed, the
centers of mass of two macromolecules can coincide without violation
of the excluded volume conditions, hence bringing about a bounded 
interaction. Moreover, the same mechanisms that exist for tuning
the usual, diverging interactions between colloidal particles can be
applied in order to tune the bounded interactions: the solvent quality,
temperature, chain length, salt concentration etc.\ will all affect
the effective potential. Thus, it appears to be useful to consider
such potentials in some generality in order to be able to draw 
conclusions about the expected phase behavior of systems interacting
by means of these.

Two model systems in this category have already
been studied in some detail.
One is the penetrable spheres model (PSM) 
\cite{marquest:witten:89,likos:pensph:98}, in which the pair potential
is a positive constant $\varepsilon$ for distances $r < \sigma$ and
vanishes otherwise. The other is the Gaussian core model (GCM), introduced
in the mid-seventies by Stillinger \cite{stillinger:76}. In the GCM, the
pair potential $v(r)$ has the form $v(r) = \varepsilon\exp[-(r/\sigma)^2]$,
with $\varepsilon$ being an energy and $\sigma$ a length scale. 
It has been shown that the GCM models very accurately the effective
interactions between the centers of mass of linear polymer 
chains \cite{olaj:lantschbauer:77,grosberg:82,schaefer:baumgaertner:86,
krueger:etal:89,dautenhahn:hall:94,ard:finken:hansen:99,ard:peter:00,bolhuis:jcp:00}.

The PSM was studied by means of cell-model calculations and
computer simulations \cite{likos:pensph:98},
liquid-state integral equation theories
\cite{fernaud:jcp:00} and density-functional
theory \cite{schmidt:cecam:99}; the fluid structure 
of the PSM has been further studied recently by 
Rosenfeld {\it et al.} by using ideas based on the
universality of the bridge functional \cite{rosenfeld:etal:pensph}.
It was found that no reentrant
melting takes place because the solid always lowers its free energy by
allowing for multiply occupied crystal sites, a mechanism that is
called {\it clustering} \cite{likos:pensph:98}. The clustering
mechanism stabilizes therefore
the solid at all temperatures. Hence, the topology of the phase diagram
of the PSM is similar to that of power-law diverging potentials,
when details about the clustering structure of the solids are disregarded.
On the other hand, the GCM has been
studied even more extensively by means of molecular dynamics
simulations \cite{stillinger:weber:78,stillinger:weber:80}, high-temperature
expansions \cite{stillinger:jcp:79} and the discovery
of exact duality relations in the crystalline state \cite{stillinger:prb:79}.
Recently, 
a full statistical-mechanical
study of the GCM was performed and it was
established that the topology of the phase diagram of
the GCM resembles that of star polymers. Freezing and reentrant melting
accompanied by an upper freezing temperature were quantitatively 
calculated \cite{lang:etal:gcm}. The question that arises, therefore,
is the following. 
Given a nonattractive and bounded pair potential which satisfies
the following requirements guaranteeing stability and the existence of
the thermodynamic limit \cite{ruelle}:
\begin{enumerate}
\renewcommand{\labelenumi}{(\roman{enumi})}
\item it is bounded;
\item it is positive definite;
\item it decays fast enough to zero at large separations, 
      so that it is integrable and
      its Fourier transform exists;
\end{enumerate}
to which topology belongs the phase diagram of the system? In this paper,
we present an exact criterion which gives an answer to this question
and show representative results for model systems which confirm its
validity. The rest of the paper is organized as follows: in section II
we present the physical arguments supporting the mean-field theory
of the models and in section III we discuss the existence of a spinodal
instability in this theory and its implications on the phase bahavior.
We present a systematic comparison between theory and simulation
in section IV and we draw the generic phase diagrams of such systems
in section V. Finally, in section VI we summarize and conclude.

\section{The model and the mean-field limit}
 
We will work with a general interaction 
$v(r) = \varepsilon\phi(r/\sigma)$ satisfying the requirements
put forward above. Here, $\varepsilon$ and
$\sigma$ are an energy and a length, respectively, and
$\phi(x)$ is some dimensionless function. The latter does {\it not} have
to be analytic, i.e., discontinuities in the potential or its
derivatives are allowed. Without loss of generality, we assume
$\phi(0) = 1$.
Let us call 
$\hat \phi(Q) = \sigma^{-3}\tilde\phi(Q)$
the dimensionless Fourier transform (FT) of the interaction.
For more concreteness (and for the purposes of
demonstration) we introduce in addition 
the family of bounded potentials $v_{\xi}(r)$ depending 
on a tunable parameter $\xi$,
\begin{equation}
v_{\xi}(r) = \varepsilon\; {{1 + e^{-\sigma/\xi}}\over
              {1 + e^{(r-\sigma)/\xi}}},
\label{fermi}
\end{equation}
where $\xi$ is a `smoothing parameter' having dimensions
of length. The case $\xi = 0$ recovers the PSM whereas as $\xi$ grows
the interaction becomes smoother. Due to its resemblance to the
Fermi-Dirac distribution, we call this family the Fermi distribution
model (FDM). The additional factor $1 + e^{-\sigma/\xi}$ in the
numerator of the rhs of eq.\ (\ref{fermi}) ensures that the potential
varies from $\varepsilon$ at $r = 0$ to zero at $r \rightarrow \infty$,
for all $\xi$.

We introduce dimensionless measures of temperature and density 
as
\begin{eqnarray}
t & = & {{k_BT}\over{\varepsilon}} = \left(\beta\varepsilon\right)^{-1};
\\
\eta & = & {{\pi}\over{6}}\rho\sigma^3  =  {{\pi}\over{6}}\bar\rho,
\label{params}
\end{eqnarray}  
where $k_B$ is Boltzmann's constant and $\rho = N/V$ is the density of
a system of $N$ particles in the volume $V$. We will refer to $\eta$
as the `packing fraction' of the system. 

The key idea for examining
the high temperature and/or high density limit of such model systems
is the following. 
We consider in general 
a spatially modulated density profile $\rho({\bf r})$ which does 
not vary too rapidly on the scale $\sigma$ set by the interaction.
At high densities, $\rho\sigma^3 \gg 1$, the average interparticle
distance $a \equiv \rho^{-1/3}$ becomes vanishingly small, 
and it holds $a \ll \sigma$, i.e., the potential is extremely
long-range. Every particle is simultaneously interacting with
an enormous number of neighboring molecules and in the absence of
short-range excluded volume interactions the excess free energy
of the system \cite{evans:79}
can be approximated by a simple mean-field term, equal to
the internal energy of the system:
\begin{equation}
F_{\rm ex}[\rho({\bf r})] \cong
{{1}\over{2}} \int\int d{\bf r} d{\bf r'} v(|{\bf r} - {\bf r'}|)
\rho({\bf r}) \rho({\bf r'}),
\label{dft.mfa}
\end{equation}
with the approximation becoming more accurate with increasing density.
Then, eq.\ (\ref{dft.mfa}) immediately implies that in this limit
the direct correlation function $c(r)$ of the system, 
defined as \cite{evans:79}
\begin{equation}
c(|{\bf r} - {\bf r'}|;\rho) =
-\lim_{\rho({\bf r}) \to \rho}
{{\delta^{2} \beta F_{\rm ex}[\rho({\bf r})]}\over
 {\delta \rho({\bf r}) \delta \rho({\bf r'})}},
\label{dcf.dft}
\end{equation}
becomes independent of the density and is simply proportional
to the interaction, namely
\begin{equation}
c(r) = -\beta v(r).
\label{msa}
\end{equation} 
Using the last equation, together with the Ornstein-Zernike 
relation \cite{hansen:mcdonald}, we readily obtain an analytic expression 
for the structure factor $S(Q)$ of the system as
\begin{equation}
S(Q) = {{1}\over{1 + \bar\rho t^{-1} \hat \phi(Q)}}.
\label{sofq.analytic}
\end{equation}
This mean-field approximation (MFA) has been put forward and examined
in detail in the context of the Gaussian core model independently
by Lang {\it et al.}\ \cite{lang:etal:gcm} and by 
Louis {\it et al.}\ \cite{ard:mft:00}. The model is particularly
relevant from the physics point of view,
due to its connection to the theory of effective interactions
between polymer chains \cite{ard:peter:00,bolhuis:jcp:00}. 
Here, we establish the validity of the MFA at high densities for
bounded, positive-definite interactions {\it in general} and we
examine its implications for the global phase behavior of such
systems.

Bounded and positive-definite interactions have been
studied in the late seventies by
Grewe and Klein \cite{grewe:klein:jmpa:77,grewe:klein:jmpb:77}.
The authors considered a slightly different model than the one
considered here, namely a Kac potential \index{Kac potential}
of the form:
\begin{equation}
v(r) = \gamma^d\psi(\gamma r),
\label{kac}
\end{equation}
where $d$ is the dimension of the space and $\gamma \geq 0$ is a
parameter controlling the range {\it and} strength of the potential.
Moreover, $\psi(x)$ is a nonnegative, bounded and integrable function:
\begin{equation}
0 \leq \psi(x) \leq A < \infty,
\qquad C = \int d^d{\bf x} \psi(x) < \infty.
\label{integrable}
\end{equation}
Grewe and Klein were able to show rigorously that at the limit
$\gamma \to 0$, the direct correlation function of a system interacting
by means of the potential (\ref{kac}) is given by eq.\ (\ref{msa})
above. The connection with the case we are discussing here is
straightforward: as there are no hard cores in the system or a
lattice constant to impose a length scale, the only relevant
length is set by the density and is
equal to $\rho^{-1/3}$ in our model and by
the parameter $\gamma^{-1}$ in model (\ref{kac}). In this respect,
the limit $\gamma \to 0$ in the Kac model of Grewe and Klein
is equivalent to the limit $\rho \to \infty$ considered here.
However, in the Kac model the strength of the interaction goes
to zero simultaneously with the increase in its range. Moreover,
the validity of the mean-field expression at large but finite
densities and at low temperatures has not been tested in detail. 

\section{Spinodal instability and freezing}
We employ the MFA as a physically motivated 
{\it working hypothesis} for now and, by direct comparison with
simulation results, we will show later that it is indeed valid.
Within the framework of this theory, an exact criterion can be formulated,
concerning the stability of the liquid phase at high temperatures and
densities.
The function $\phi(x)$ 
was assumed to be decaying monotonically
from unity at $x = 0$ to zero at 
$x \rightarrow \infty$. For the function
$\hat\phi(Q)$, there are two possibilities:
(i) It has a monotonic decay from the 
value $\hat\phi(Q = 0) = \sigma^{-3}\int d{\bf x} \phi(x) > 0$ to the value 
$\hat\phi(Q) = 0$ at $Q \rightarrow \infty$. We call such potentials
$Q^{+}$-{\it potentials}. Obviously, the Gaussian interaction 
belongs to this class. 
(ii) It has oscillatory
behavior at large $Q$, with the implication that it is a nonmonotonic
function of $Q$, attaining necessarily negative values for certain ranges
of the wavenumber. We call such potentials $Q^{\pm}$-{\it potentials}. 
Long-range oscillations in $Q$-space imply that $\phi(x)$
changes more rapidly from unity at $r = 0$ to zero at $r \rightarrow \infty$
in the $Q^{\pm}$-class than in the $Q^{+}$ one.
Moreover,
let us call $Q_{*}$ the value of $Q$ at which $\hat\phi(Q)$
attains its minimum, {\it negative} value.

If we are dealing with a $Q^{\pm}$-potential
eq.\ (\ref{sofq.analytic})
implies that $S(Q)$
has a maximum at precisely the wavevector $Q_*$ where $\hat\phi(Q)$
attains its negative minimum, $-|\hat\phi(Q_*)|$ and this maximum
becomes a {\it singularity} at
the `spinodal line' $\bar\rho t^{-1}|\hat\phi(Q_*)| = 1$, signaling the so-called Kirkwood instability of the
system \cite{grewe:klein:jmpb:77,klein:grewe:jcp:80,klein:brown:jcp:81,klein:etal:physica:94}. 
The theory
has a divergence, implying that the underlying assumption of a uniform
liquid is not valid and the system must reach a crystalline state.
Indeed, on the basis of the fluctuation-dissipation
theorem, $S(Q)$ can be interpreted as a response function of the
density to an infinitesimal external modulating field at
wavenumber $Q$ \cite{hansen:mcdonald}
and a diverging value of this response function 
clearly signals an instability.
If the Fourier transform of $\phi(x)$ has negative Fourier components,
then an increase in temperature can be compensated by an increase
in density in the denominator of eq.\ (\ref{sofq.analytic}), so that
$S(Q_*)$ will have a divergence at all $t$. We thus conclude
that $Q^{\pm}$-{\it systems freeze at all temperatures}.

If we are dealing with a $Q^{+}$-potential ($\hat\phi(Q)$ monotonic), then
eq.\ (\ref{sofq.analytic}) implies that $S(Q)$ is also
a monotonic function of $Q$ at high densities \cite{lang:etal:gcm}.
For such
potentials, one can always find a temperature high enough, so that
the assumptions of eq.\ (\ref{dft.mfa}) hold and then
eq.\ (\ref{sofq.analytic}) forces the conclusion that freezing of the
system is impossible at such temperatures. This does not imply, of
course, that such systems do not freeze at all; one simply has to go
to a low enough temperature and density, 
so that the mean-field assumption does
not hold and the interaction is much larger that the thermal
energy. Then, the system will display a hard-sphere type of freezing,
to be discussed more explicitly below. 
An upper freezing temperature $t_{\rm u}$ must exist for $Q^{+}$-potentials,
implying that such systems must remelt at $t < t_{\rm u}$
upon increase of the density. Hence, we reach the conclusion that
$Q^{+}$-{\it systems display an upper freezing temperature and reentrant
melting}. The criterion says nothing about the crystal structure of
the solid, however, which always depends on the details of the interaction
as well as the density \cite{lang:etal:gcm,stillinger:stillinger:97}.

For potentials is in the $Q^{+}$-class, the mean-field arguments
presented above hold not only at high temperatures but also at low ones,
provided that the requirement $\rho\sigma^3 \gg 1$ is satisfied, 
because these are molten at high densities for 
all nonzero temperatures. The validity of the mean-field theory
for $Q^{+}$-type systems, even at very low temperatures,
was confirmed recently by direct comparison
with simulation results for the particular case of the Gaussian 
potential \cite{lang:etal:gcm}. If the potential is in the
$Q^{\pm}$-class, the mean field approximation holds provided that
the system is not already frozen, as we will confirm shortly.
Moreover, both kinds of systems
display an unusual kind of `high density ideal gas' limit. Indeed,
taking the expression (\ref{sofq.analytic}) for $S(Q)$ and using
the relation $S(Q) = 1 + \bar\rho\hat h(Q)$ \cite{hansen:mcdonald},
where $\hat h(Q)$ is the dimensionless Fourier transform 
of the pair correlation
function $h(r)$ of the uniform fluid, we obtain:
\begin{equation}
\hat h(Q) = -{{t^{-1} \hat\phi(Q)}\over
                {1 + \bar\rho t^{-1}\hat \phi(Q)}}.
\label{hofq}
\end{equation}

At low $Q$'s, where $\hat\phi(Q)$ is of order unity, the term proportional
to the density in the denominator dominates in the limit of high
densities and $\hat h(Q)$ scales as $-1/\bar\rho \rightarrow 0$. At high
$Q$'s, the Fourier transform $\hat \phi(Q)$ in the numerator is itself small,
with the result that $\hat h(Q)$, and hence also the correlation
function $h(r)$, is approaching zero. This, in turn, means that the
radial distribution function $g(r) = h(r) + 1$ is very close to unity
in this limit and deprived of any significant structure
for all values of $r$ and it only has some small structure at small $r$,
which is in fact more pronounced for $Q^{\pm}$-potentials than for
$Q^{+}$ ones. 
In this limit, the hypernetted chain (HNC) 
closure becomes exact, as the exact relation
$g(r) = \exp[\beta v(r) + h(r) - c(r) -B(r)]$, combined with the 
limits $g(r) \rightarrow 1$, $h(r) \rightarrow 0$ and 
$c(r) \rightarrow -\beta v(r)$ forces the bridge function $B(r)$ to
vanish. Moreover, 
eqs.\ (\ref{sofq.analytic}) and (\ref{hofq}) reveal that
the systems obey a scaling law, namely that the functions $S(Q)$ and
$th(r)$ do not depend on $\bar\rho$ and $t$ separately but only on
the ratio $\bar\rho/t$.  

Systems in the $Q^{\pm}$-class freeze
before the spinodal is reached. In order to make quantitative predictions,
we invoke the empirical Hansen-Verlet freezing
criterion \cite{hansen:verlet:69,hansen:schiff:73}, which states that
a system crystallizes when $S(Q)$ at its main peak attains,
approximately, the value $S(Q_{*}) = S_{\rm m} \cong 3$. 
Although this criterion
was originally put forward for hard, atomic interactions 
(HS, Lennard-Jones etc.), recent detailed analyses have demonstrated
that it holds for the freezing and the remelting transitions of
ultrasoft particles such as star polymers \cite{watzlawek:etal:prl:99,watzlawek:etal:jpcm:98} and even for the nondiverging Gaussian 
interaction \cite{lang:etal:gcm}. Hence, we assume that it is valid 
for the general class of systems we consider here and 
combining it
with eq.\ (\ref{sofq.analytic}), we obtain the
equation of the freezing line $t_{\rm f}(\eta)$ as
\begin{equation}
t_{\rm f}(\eta) = \frac{6|\hat\phi(Q_*)|}{\pi(1 - S_{\rm m}^{-1})}\;\eta
              \cong 2.864\;|\hat\phi(Q_*)|\eta.
\label{freezing.line}
\end{equation}
The
value $|\hat \phi(Q_*)|$ determines the slope of the freezing line
at the high $(t,\eta)$ part of the phase diagram.

\section{Comparison with simulations}

We now wish to put these arguments in a strong test, using 
the concrete family of the 
FDM, given by eq.\ (\ref{fermi}). 
First of all, we have calculated
the Fourier transform of the potential $v_{\xi}(r)$ of eq.\ (\ref{fermi})
numerically, establishing that members of the FDM with 
$\xi < \xi_c$ belong to the $Q^{\pm}$ class and members with
$\xi > \xi_c = 0.49697$ to the $Q^{+}$ one. The GCM is also a member
of the latter class that we will discuss in what follows.

\subsection{Systems displaying clustering}

As examples of systems displaying clustering transitions,
we have taken the extreme (and by now well-studied case)
case $\xi = 0$ (the PSM) as well as the case $\xi = 0.1$ of the
Fermi distribution model of eq.\ (\ref{fermi}). We have
performed standard
Monte Carlo (MC) $NVT$ simulations for a large number of values
for the temperature and
density. We begin with the PSM for which
the analytical expression (\ref{sofq.analytic}) takes
the form
\begin{equation}
S(Q) = \left[1+24\eta t^{-1}\left({{\sin(Q\sigma)-(Q\sigma)\cos(Q\sigma)}
                            \over{(Q\sigma)^3}}\right)\right]^{-1}.
\label{psm.analytic}
\end{equation}

\begin{figure}[hbt]
   \begin{center}
   \begin{minipage}[t]{8.0cm}
         \includegraphics[width=8.0cm,height=6.0cm]{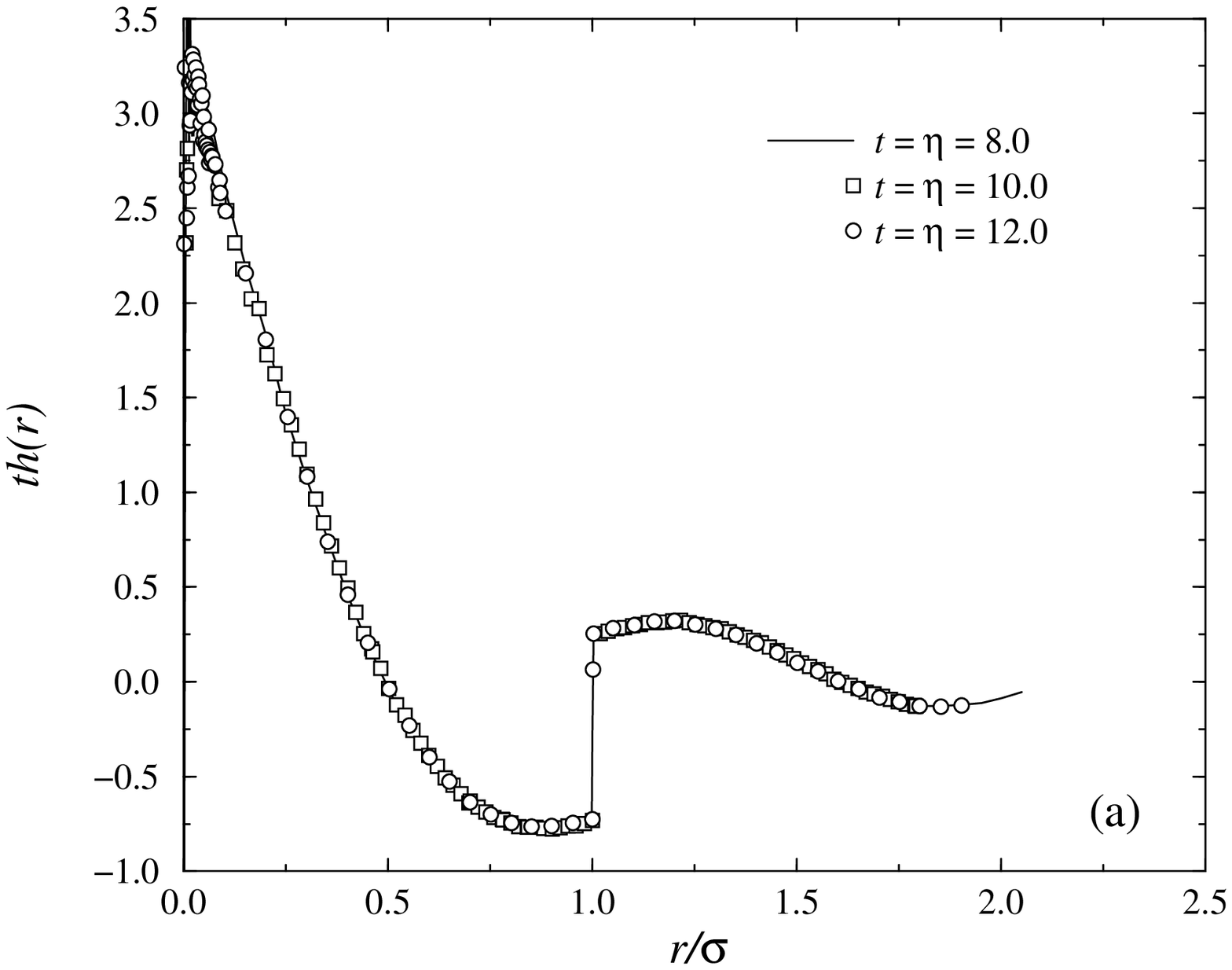}
   \end{minipage}
   \begin{minipage}[t]{8.0cm}
         \includegraphics[width=8.0cm,height=6.0cm]{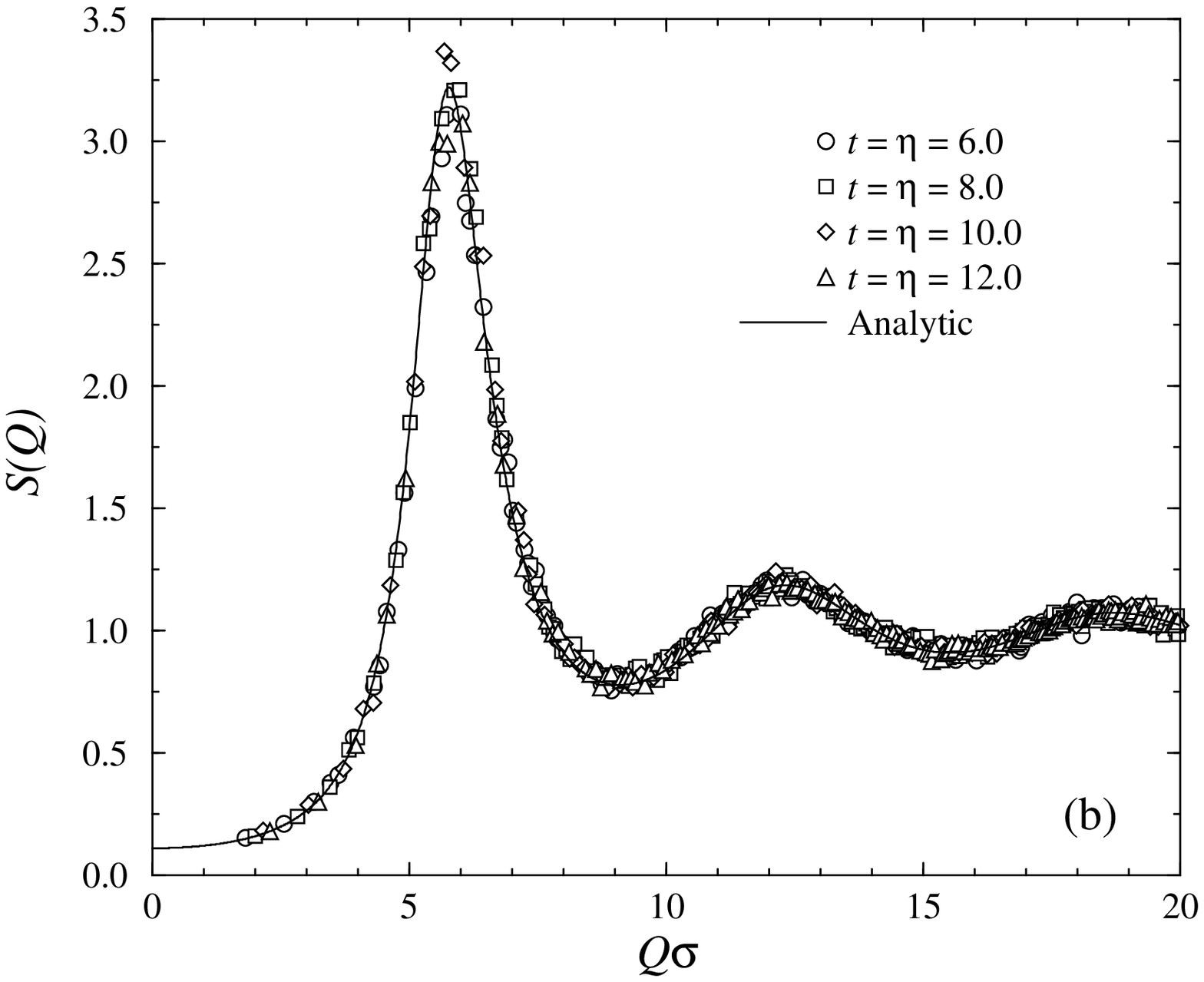}
   \end{minipage}
   \end{center}
   \caption{(a) The product $th(r)$
            for a FDM with $\xi = 0$ (PSM), along
            the diagonal $t = \eta$ at high packing fractions,
            as obtained from MC simulations. The results close to
            $r = 0$ are noisy due to poor statistics there.
            All results collapse
            onto a single curve. (b) The corresponding structure
            factors $S(Q)$, shown together with the analytical
            result of eq.\ (\ref{psm.analytic}).}
\label{psm.diag.plot}
\end{figure}

\begin{figure}[hbt]
   \begin{center}
   \begin{minipage}[t]{8.0cm}
         \includegraphics[width=8.0cm,height=6.0cm]{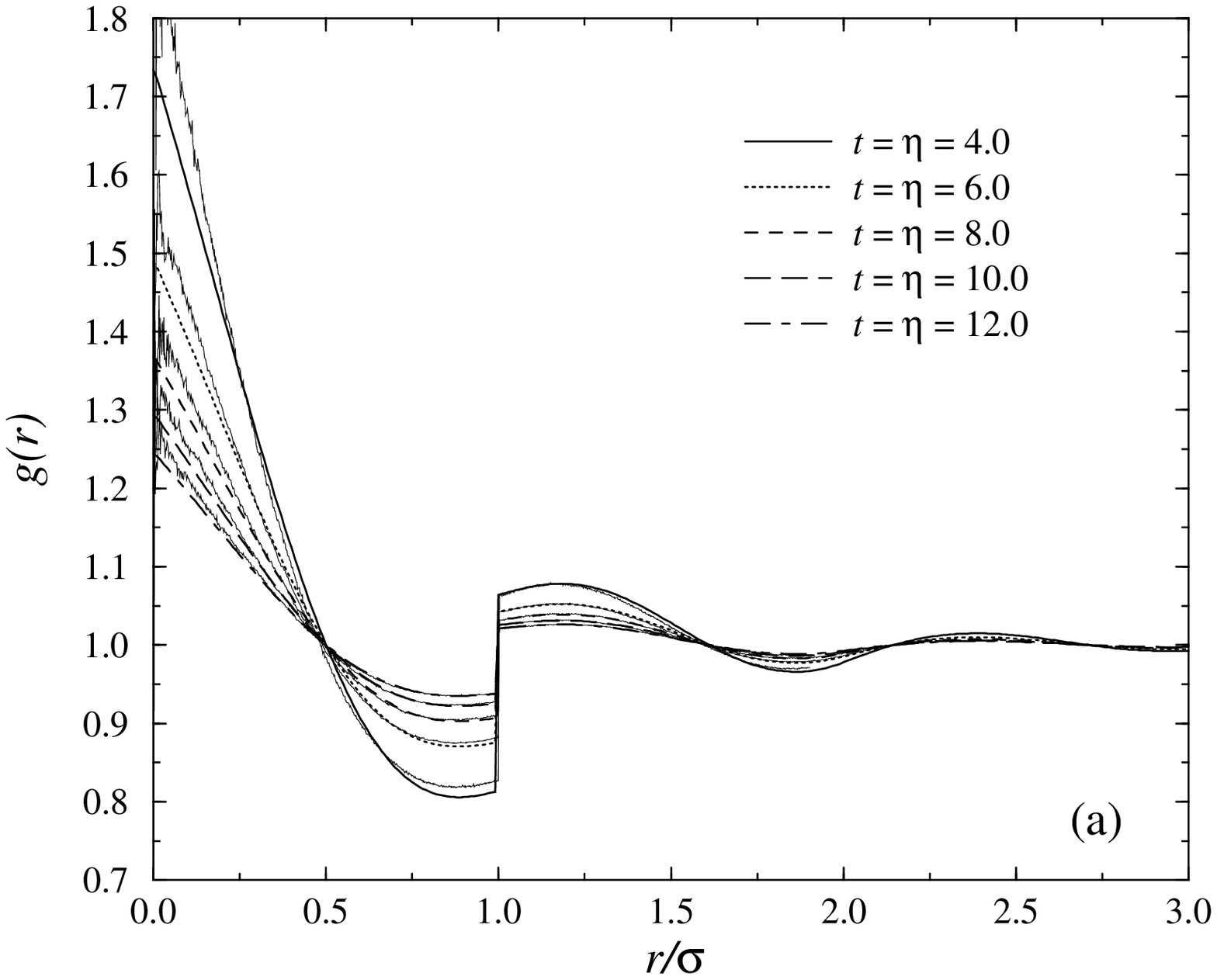}
   \end{minipage}
   \begin{minipage}[t]{8.0cm}
         \includegraphics[width=8.0cm,height=6.0cm]{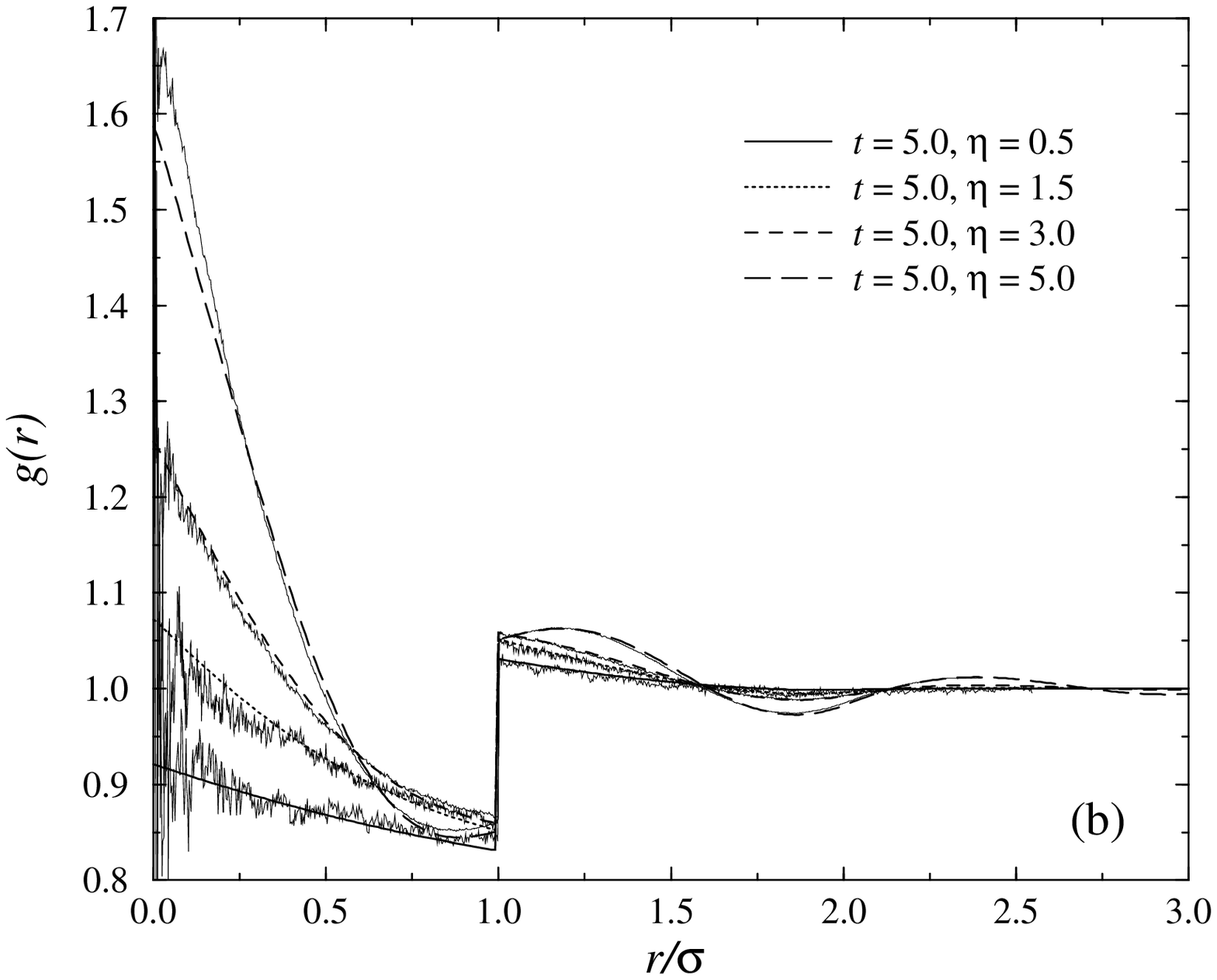}
   \end{minipage}
   \end{center}
   \caption{(a) The function $g(r)$ of the PSM for selected points
            along the `diagonal' $t = \eta$ as obtained from theory
            (thick lines) and simulation (thin lines). (b) Same
            but now for fixed temperature $t = 5$ and increasing
            packing fraction $\eta$.}
\label{psm.gofrI.plot}
\end{figure}

\begin{figure}[hbt]
   \begin{center}
   \begin{minipage}[t]{8.0cm}
         \includegraphics[width=8.0cm,height=6.0cm]{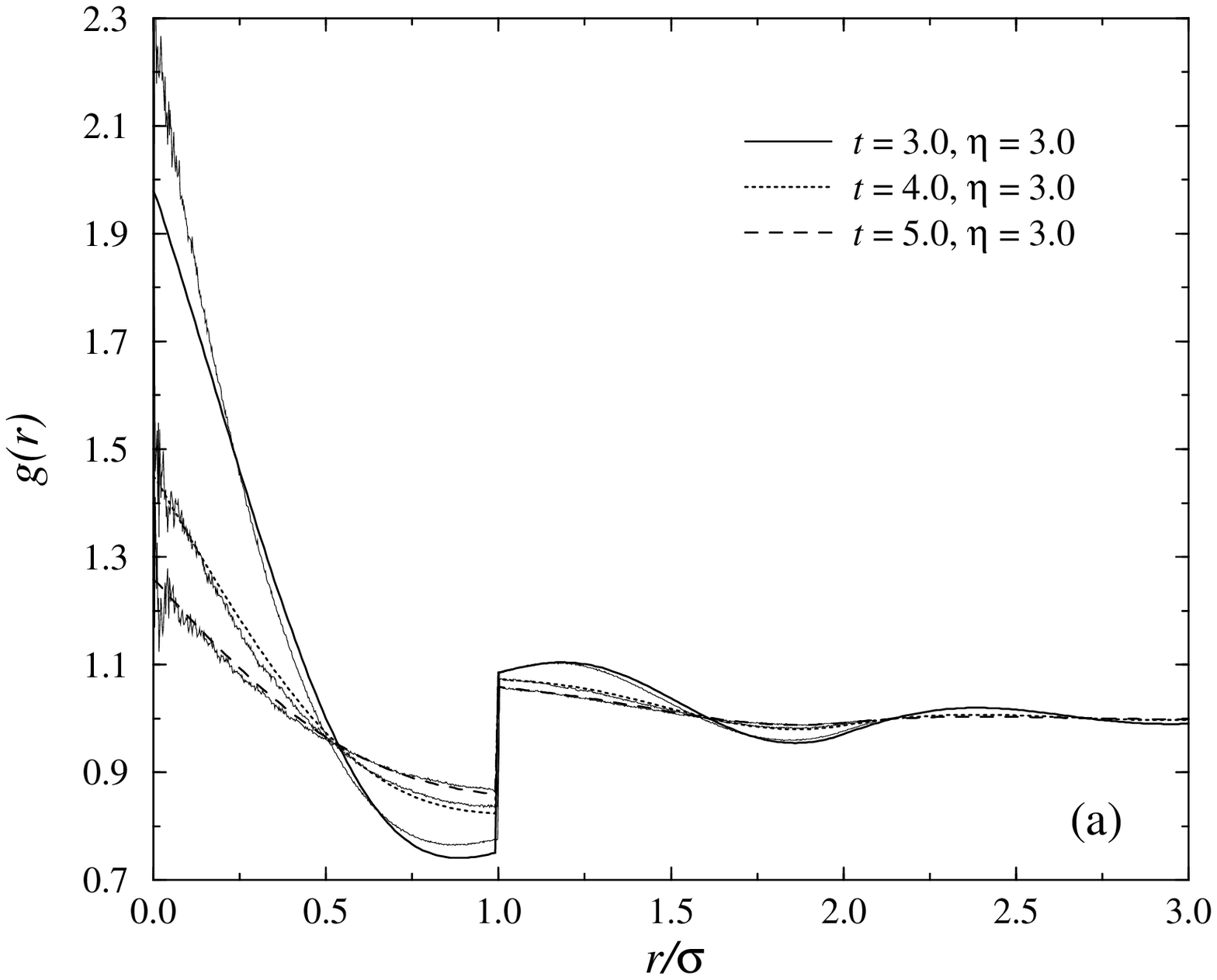}
   \end{minipage}
   \begin{minipage}[t]{8.0cm}
         \includegraphics[width=8.0cm,height=6.0cm]{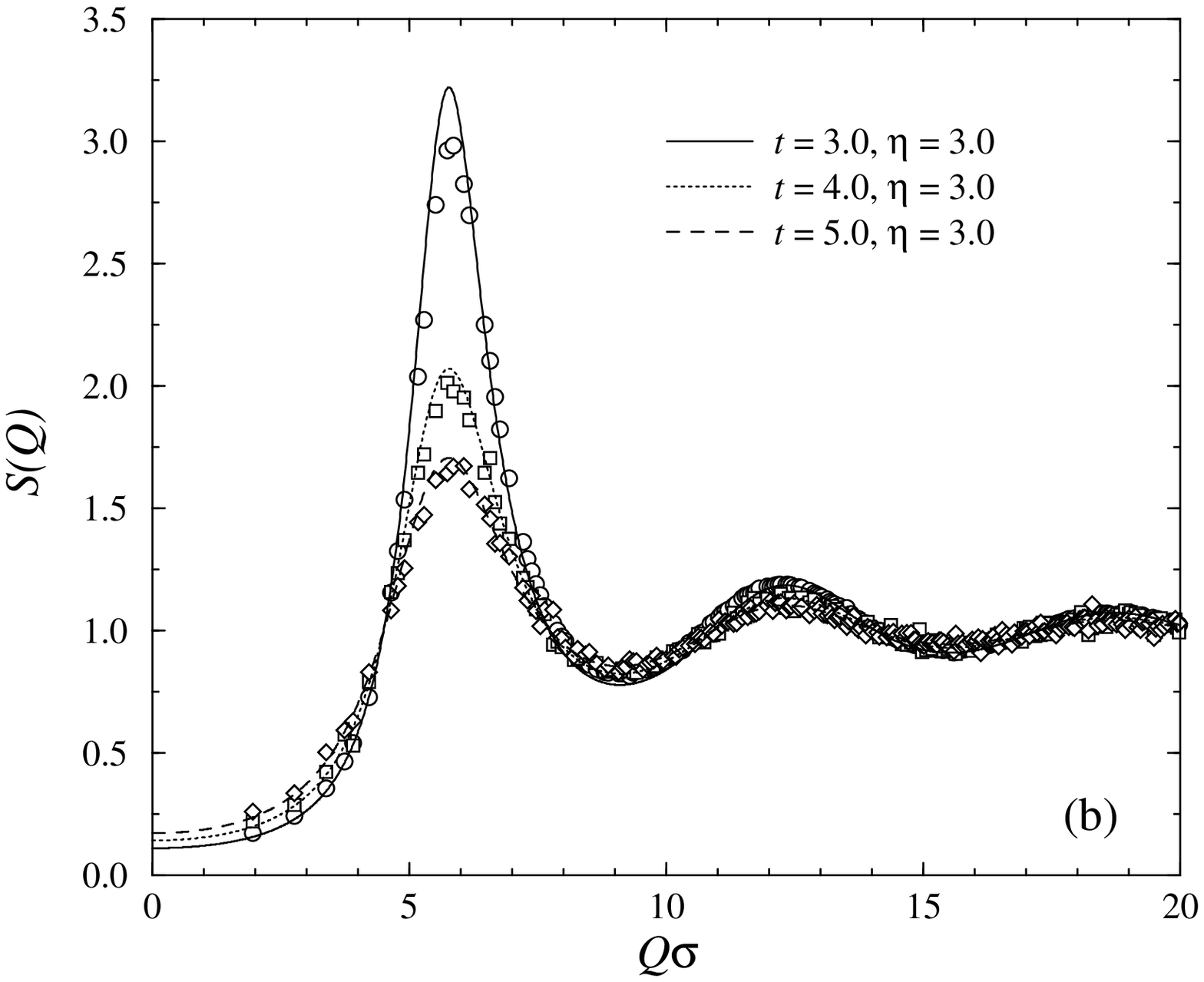}
   \end{minipage}
   \end{center}
   \caption{(a) Same as Fig.\ \ref{psm.gofrI.plot}(b) but now for
            fixed $\eta = 3.0$ and increasing temperature. (b)
            The structure factors at the thermodynamic points
            of (a), comparison between theory (lines) and simulation
            (points).}
\label{psm.gofrII.plot}
\end{figure}

The
high temperature-high density freezing line of eq.\ (\ref{freezing.line})
takes for this choice of $\xi$ the form $t_{\rm f}(\eta) = 1.033\eta$.
To test the analytical expression of eq.\ (\ref{psm.analytic}), we
move along the `diagonal' $t = \eta$,
a combination that lies almost on the Hansen-Verlet estimate for 
the location of the freezing line.
In Fig.\ \ref{psm.diag.plot} we show the comparison of the
analytical results with those obtained from the MC simulations for
$S(Q)$ and we also demonstrate that the MC curves 
for the quantity $th(r)$ all collapse 
onto a single line, amply demonstrating the validity of the 
mean-field approximation for the PSM. 
In order to further investigate the validity of the MFA, we have
performed MC simulations in a variety of thermodynamic points
and we present a selection of the obtained results. We present
a selection of these in Figs.\ \ref{psm.gofrI.plot} and
\ref{psm.gofrII.plot} and discuss them below.

In Fig.\ \ref{psm.gofrI.plot}(a) we show a comparison between the
MC and MFA results for the radial distribution function $g(r)$ along
the `diagonal' $t = \eta$. It can be seen that the agreement between the
two is already very good at $t = \eta = 4.0$ and thereafter it improves
markedly with increasing temperature and density. The results obtained from
the present theory are of the same quality as those obtained
by Fernaud {\it et al.} \cite{fernaud:jcp:00}, who used the sophisticated
zero-separation (ZSEP) closure to investigate the liquid structure
of the system. This closure involves three self-consistency parameters,
determined in such a way that the virial-compressiblity,
Gibbs-Duhem and zero-separation consistency conditions are fulfilled.
At the same time, the present results are of the same quality as
the recently obtained results of
Rosenfeld {\it et al.} \cite{rosenfeld:etal:pensph}, based on ideas
of the universality of the bridge functional. 

In Fig.\ \ref{psm.gofrI.plot}(b) we perform the same comparison but
now at {\it fixed temperature} $t = 5.0$ and increasing packing fraction
$\eta$. As can be seen, at this temperature, the MFA, which was 
originally formulated as a high-density approximation, proves to
perform extremely well even at intermediate packings, $\eta = 0.5$
for instance. This is a direct consequence of the boundedness of
the interaction combined with a temperature $t \gg \varepsilon$.
Indeed, for small densities, the direct correlation function
tends to the Mayer function, 
$c(r) \cong \exp\left[-\beta v(r)\right] - 1$ \cite{hansen:mcdonald}.
If we are dealing
with a bounded interaction at high temperature, we can linearize the
exponential, obtaining $c(r) \cong -\beta v(r)$ at low densities, which
matches with the MFA expression, eq.\ (\ref{msa}), at high densities,
thus leading to the conclusion that the MFA is an excellent approximation
at {\it all densities}. For unbounded interactions the linearization
of the exponential is evidently impossible.

In Fig.\ \ref{psm.gofrII.plot} we present a comparison between MC
and MFA at {\it fixed packing fraction} $\eta = 3.0$ and increasing
temperature. As can be clearly seen, the validity of the MFA improves
with increasing temperature. For bounded interactions, an increasing
temperature implies a `washing-out' of the correlation effects 
caused by the (increasingly weak) interaction effects and a tendency
of the system towards the particular `high-density ideal gas' limit
characterized by the tendency of the function $g(r)$ towards unity.
However, it is an interesting peculiarity of these systems that
unlike the usual ideal gas, the limit $g(r) \to 1$ (or, equivalently,
$h(r) \to 0$) does {\it not} imply a corresponding limit $S(Q) \to 1$.
Though the Fourier transform of $h(r)$, $\hat h(Q)$, tends to zero
as $\rho^{-1}$, this is compensated by the large density $\rho$,
so that the structure factor $S(Q) = 1 + \rho\hat h(Q)$ 
displays the signature of strong ordering
through the pronounced peaks seen in Figs.\ \ref{psm.diag.plot}(b)
and \ref{psm.gofrII.plot}(b). 
    
\begin{figure}[hbt]
\begin{center}
\includegraphics[width=8.0cm,height=6.0cm]{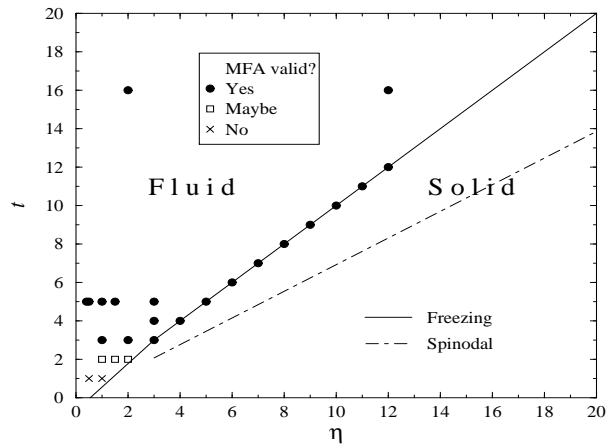}
\end{center}
\caption{The phase diagram of the PSM, along with the points where
         the mean field theory brings excellent agreement with simulation
         (filled circles), fairly good agreement (empty squares) and
         no good agreement ($\times$-symbols). These symbols should help
         delineate the domain of validity of the mean-field theory.}
\label{psm.phdg.plot}
\end{figure}

Further, we performed
MC simulations at selected points deeply inside the region
$t < t_{\rm f}(\eta)$, finding that the obtained
structure factors displayed Bragg peaks and hence confirming the
prediction that the system is frozen there. Putting all our 
results together, we draw in Fig.\ \ref{psm.phdg.plot} a 
semi-quantitative phase diagram of the PSM, accompanied by 
an assessment of the validity of the MFA at selected thermodynamic points.
The MFA appears to be an excellent approximation at all densities
above the temperature $t = 3.0$. Hence, we take 
as an estimate for the freezing line above $t = 3.0$ the MFA-Hansen-Verlet
line $t_{\rm f} = 1.033\eta \cong \eta$; for lower temperatures, we simply
connect the point $(\eta, t) = (3.0, 3.0)$ with the point
$(\eta, t) = (0.5, 0)$, which obtains from the consideration that
at $t = 0$ the PSM reduces to the hard sphere system which is
known to freeze at a fluid density $\eta_{\rm HS} \cong 0.5$. The monotonic
shape of the freezing curve for low temperatures arises from 
detailed considerations there, which can be found in 
Ref.\ \cite{likos:pensph:98}.

\begin{figure}[hbt]
   \begin{center}
   \begin{minipage}[t]{8.0cm}
         \includegraphics[width=8.0cm,height=6.0cm]{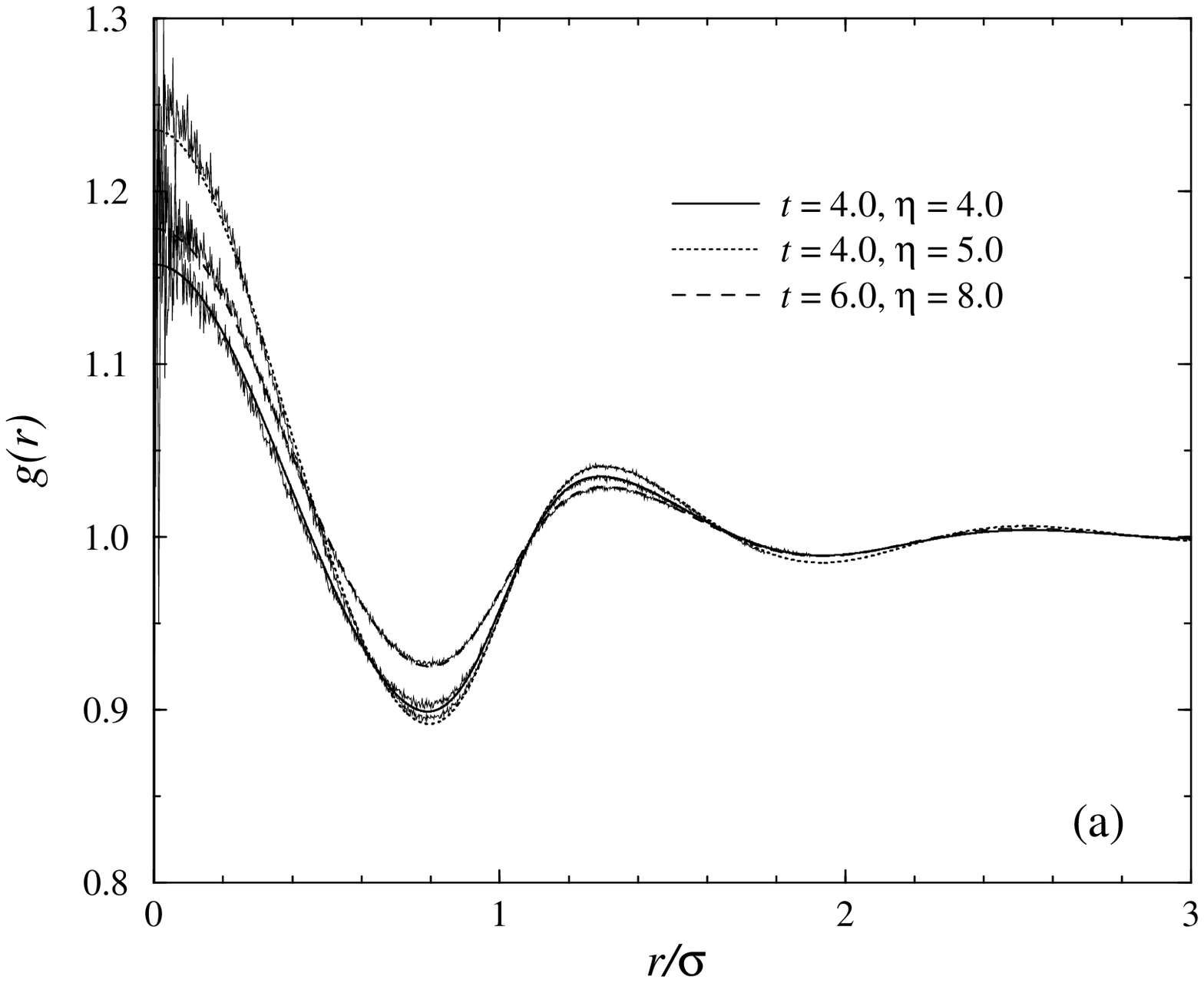}
   \end{minipage}
   \begin{minipage}[t]{8.0cm}
         \includegraphics[width=8.0cm,height=6.0cm]{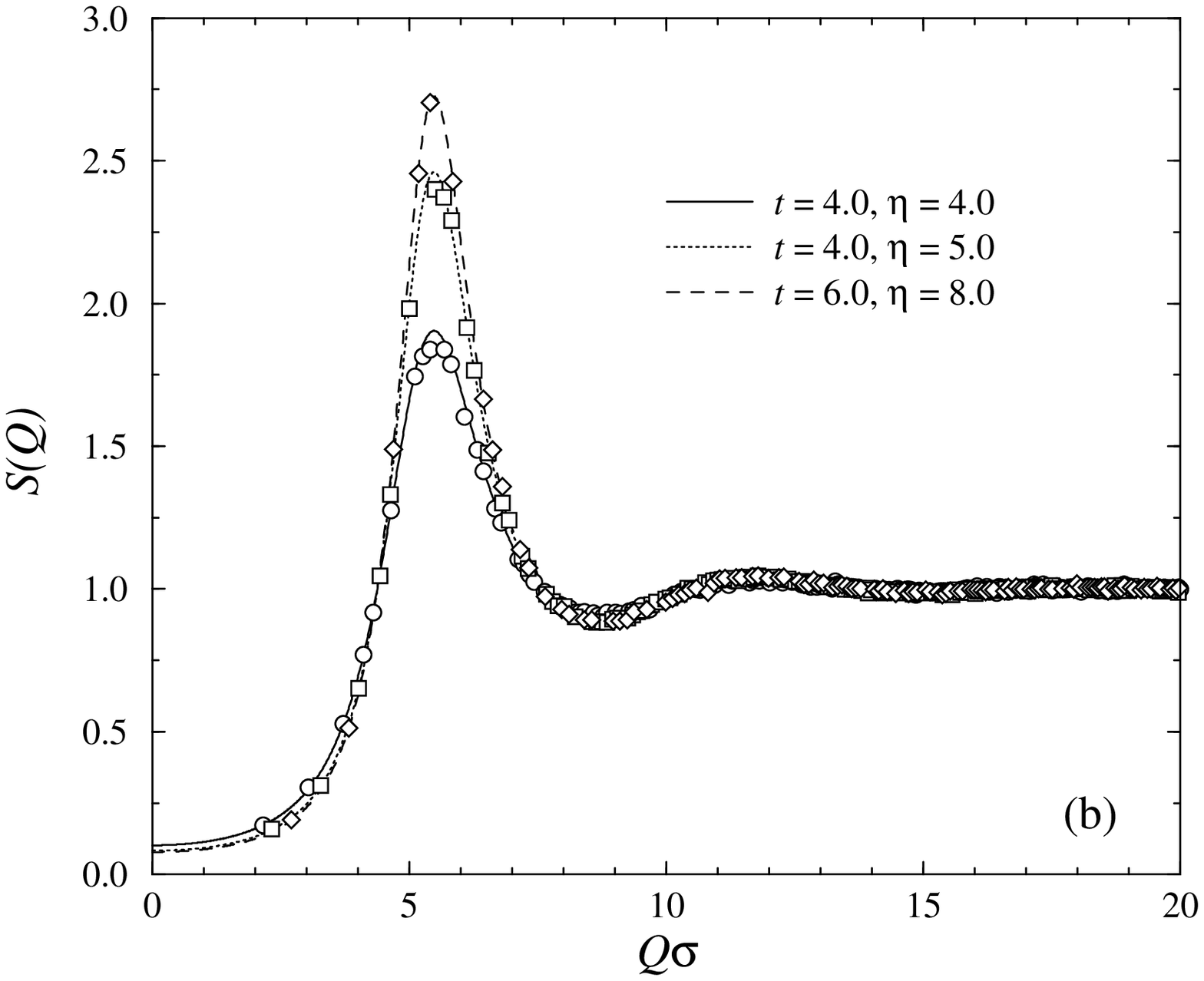}
   \end{minipage}
   \end{center}
   \caption{(a) Comparison between theory (thick lines) and simulation
            (thin lines) results for $g(r)$ of a FDM system with
            $\xi = 0.1$. (b) Comparison for the structure factors
            (lines: theory; points: simulation) for the same system
            at the thermodynamic points of (a).}
\label{xi0.1.plot}
\end{figure}

Next we present in Fig.\ \ref{xi0.1.plot}
a comparison for the FDM with $\xi = 0.1$. 
For this choice of $\xi$, the Hansen-Verlet-based
freezing line takes the form $t_{\rm f} = 0.712 \eta$.
The selected points lie in the fluid region and the
comparison indicates once more the
excellent accuracy of the MFA both for $g(r)$ and for $S(Q)$. 
The radial distribution function $g(r)$ of this model is deprived of
the jump at $r = \sigma$ seen in the PSM; the latter is caused by
the discontinuity of the PSM potential there. However, a similarity
between the $g(r)$'s of the $\xi = 0$ and $\xi = 0.1$ models is
that they both attain their maximum values at full overlaps between
the particles $r = 0$ and thereafter they decay rapidly, featuring
a depletion region around $r \approx \sigma$. This is a characteristic
pointing to a strong {\it clustering} property in the fluid
phase, a property thereafter inherited by the incipient 
thermodynamically stable crystal; the number of particles
`sitting on top of each other' and thereby occupying the same
crystal site scales linearly with density. 
In order to corroborate this claim, we can argue in two different
ways, using the liquid as a reference point.

\begin{figure}
\begin{center}
\includegraphics[width=8.0cm,height=6.0cm]{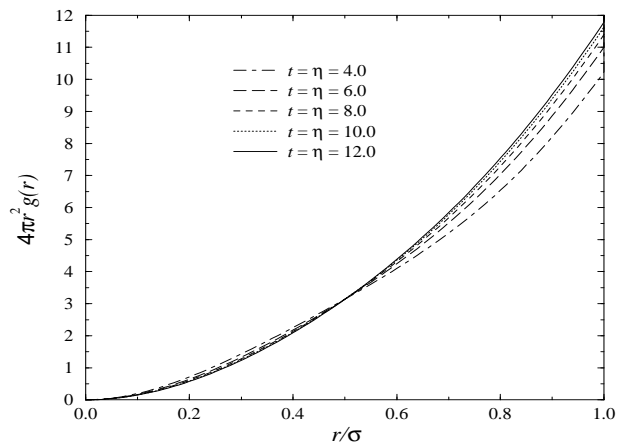}
\end{center}
\caption{The quantity $4\pi r^2 g(r)$ within the diameter $\sigma$
         of the PSM along the freezing line $t = \eta$. All the
         curves converge to a single one at high densities,
         indicating that the integral $4\pi\rho\int_0^{\sigma}
         r^2 g(r) dr$ scales linearly with density.}
\label{cluster.plot}
\end{figure}

First, let us consider the number of particles $N_{\rm c}$ in
the fluid phase whose centers are, on average, within a distance
$\sigma$ from a given particle. The number $N_{\rm c}$ is given
by the formula:
\begin{equation}
N_{\rm c} = 1 + 4\pi\rho\int_0^{\sigma} r^2 g(r) dr.
\label{cluster}
\end{equation}
In Fig.\ \ref{cluster.plot} we show the function $4\pi r^2 g(r)$ within
a particle diameter $\sigma$ for a sequence of points along the
freezing line of the PSM. As all these curves tend to a common limit
with increasing density, the integral $4\pi\int_0^{\sigma} r^2 g(r) dr$
tends to a constant and hence $N_{\rm c} \propto \rho$ at high
densities, where the second term on the rhs of eq.\ (\ref{cluster})
dominates.
  
Second, we can use the wavevector $Q_{*}$ at which the fluid structure
factor has a maximum in order to estimate for the lattice constant $a$
of the incipient crystal through the relation $a \propto Q_{*}^{-1}$.
For the models at hand, this maximum is entirely determined
by the pair potential; unlike in usual fluids featuring
diverging interactions, for which $Q_{*}$ scales as 
$\rho^{1/3}$, in our case $Q_{*}$ knows nothing about the density. Thus,
all post-freezing crystals have the same lattice constant, although
their average density is a linear function of the temperature. This
clearly shows that clustering must take place in the crystal: by allowing
more and more particles to occupy the same lattice sites, a practically
constant {\it effective density} of clusters is maintained in the
crystal, thus leading to a density-independent lattice constant.
   
\subsection{Systems displaying reentrant melting}

\begin{figure}[hbt]
   \begin{center}
   \begin{minipage}[t]{8.0cm}
         \includegraphics[width=8.0cm,height=6.0cm]{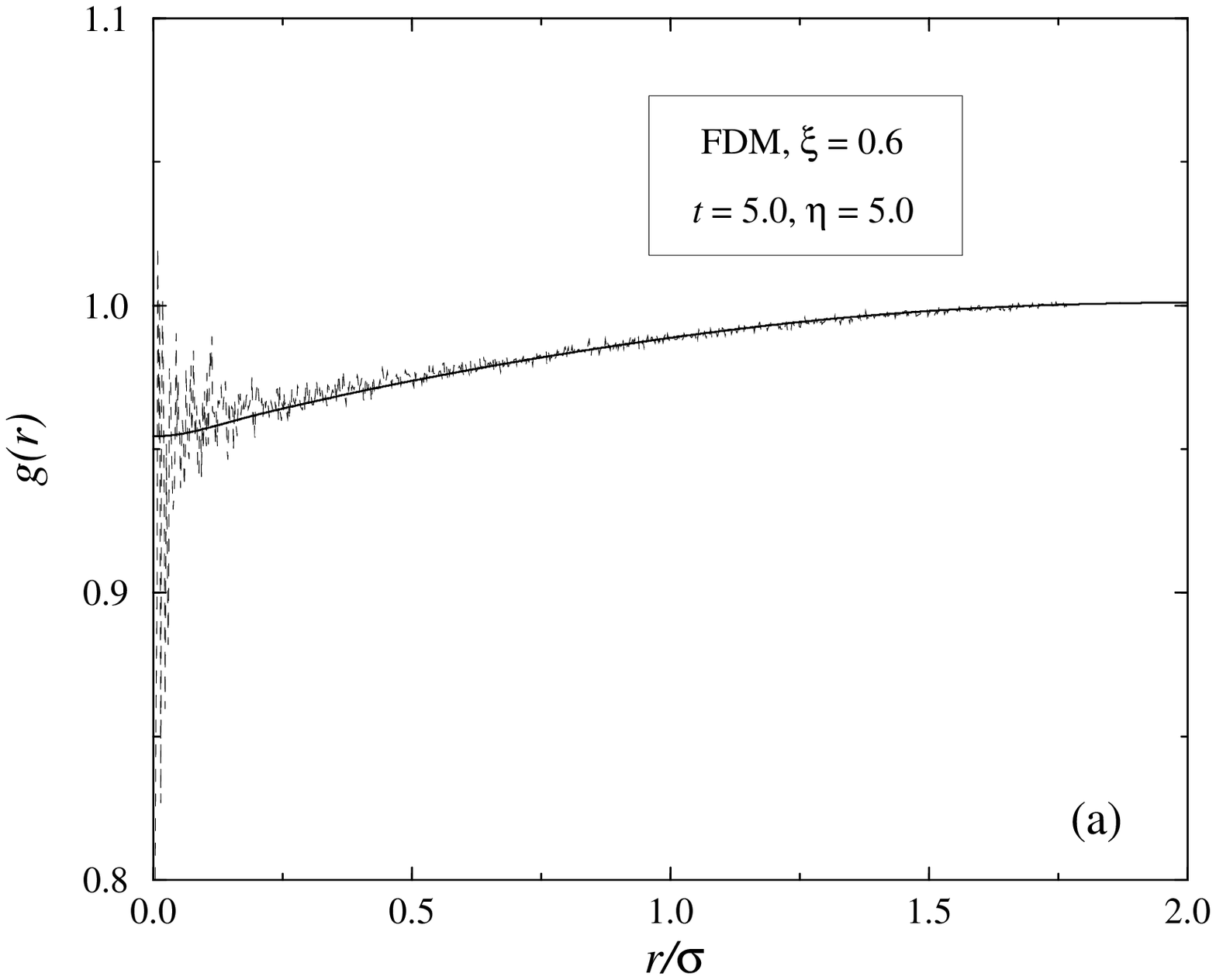}
   \end{minipage}
   \begin{minipage}[t]{8.0cm}
         \includegraphics[width=8.0cm,height=6.0cm]{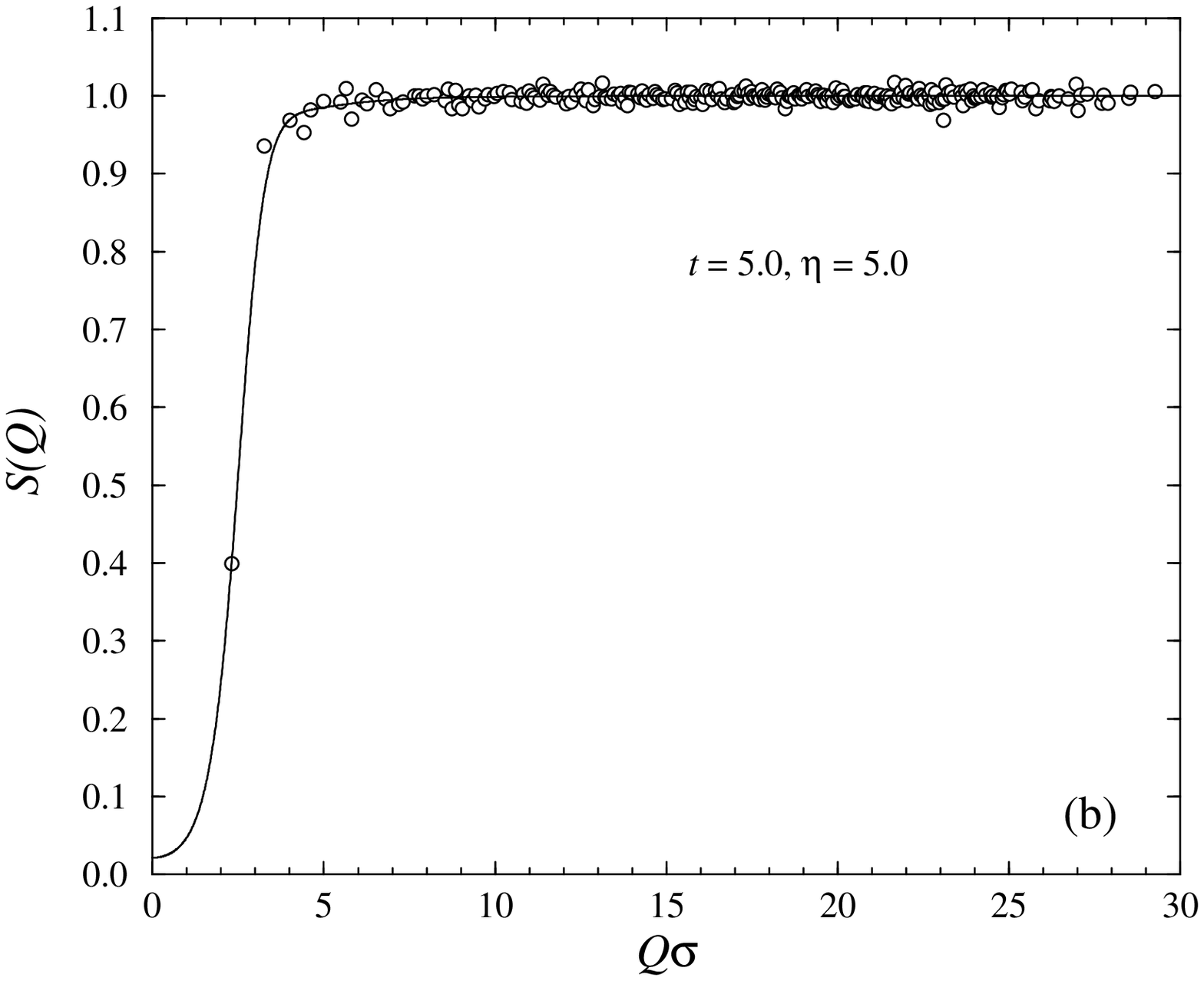}
   \end{minipage}
   \end{center}
   \caption{(a) Comparison between theory (thick lines) and simulation
            (thin lines) results for $g(r)$ of a FDM system with
            $\xi = 0.6$. (b) Comparison for the structure factor $S(Q)$
            (lines: theory; points: simulation) for the same system
            at the thermodynamic point of (a).}
\label{xi0.6.plot}
\end{figure}

We now turn our attention to the opposite case, namely pair
potentials belonging to the $Q^{+}$-class. As an example within
the FDM family, we have taken the model with parameter $\xi = 0.6$
and performed a comparison between MC and MFA results. 
A characteristic example is shown in Fig.\ \ref{xi0.6.plot}.
As can be seen in Fig.\ \ref{xi0.6.plot}(a), unlike the case
of $Q^{\pm}$-class potentials, the radial distribution function
is completely deprived of any structure, although the thermodynamic
parameters are in the same regime as those presented in 
Figs.\ \ref{psm.gofrI.plot}, \ref{psm.gofrII.plot} and
\ref{xi0.1.plot}. In fact, in the present case, $g(r)$ has
a minimum at $r = 0$, not a maximum. This complete lack of
structure is reflected in the shape of $S(Q)$, shown in 
Fig.\ \ref{xi0.6.plot}(b).

\begin{figure}[hbt]
   \begin{center}
   \begin{minipage}[t]{8.0cm}
         \includegraphics[width=8.0cm,height=6.0cm]{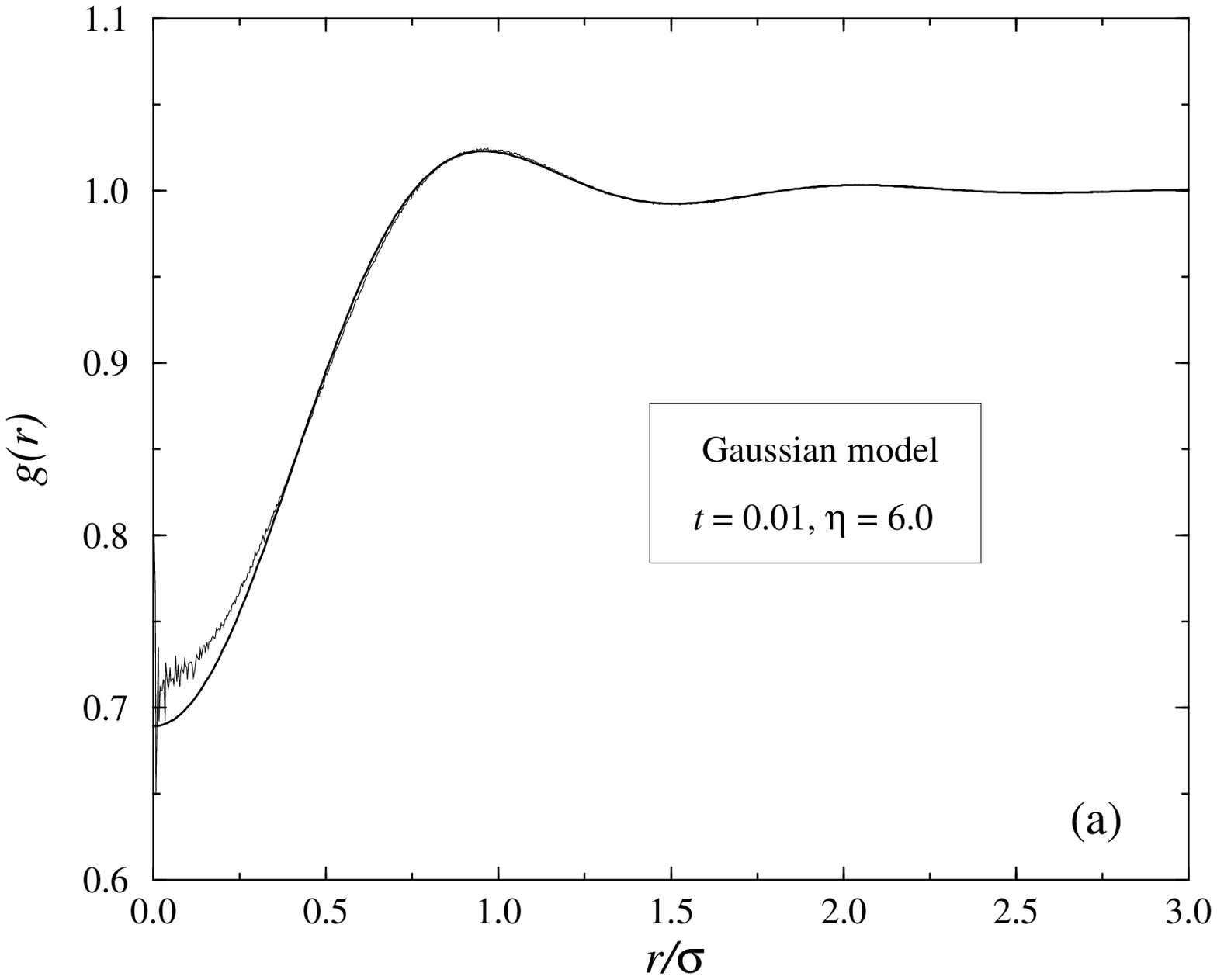}
   \end{minipage}
   \begin{minipage}[t]{8.0cm}
         \includegraphics[width=8.0cm,height=6.0cm]{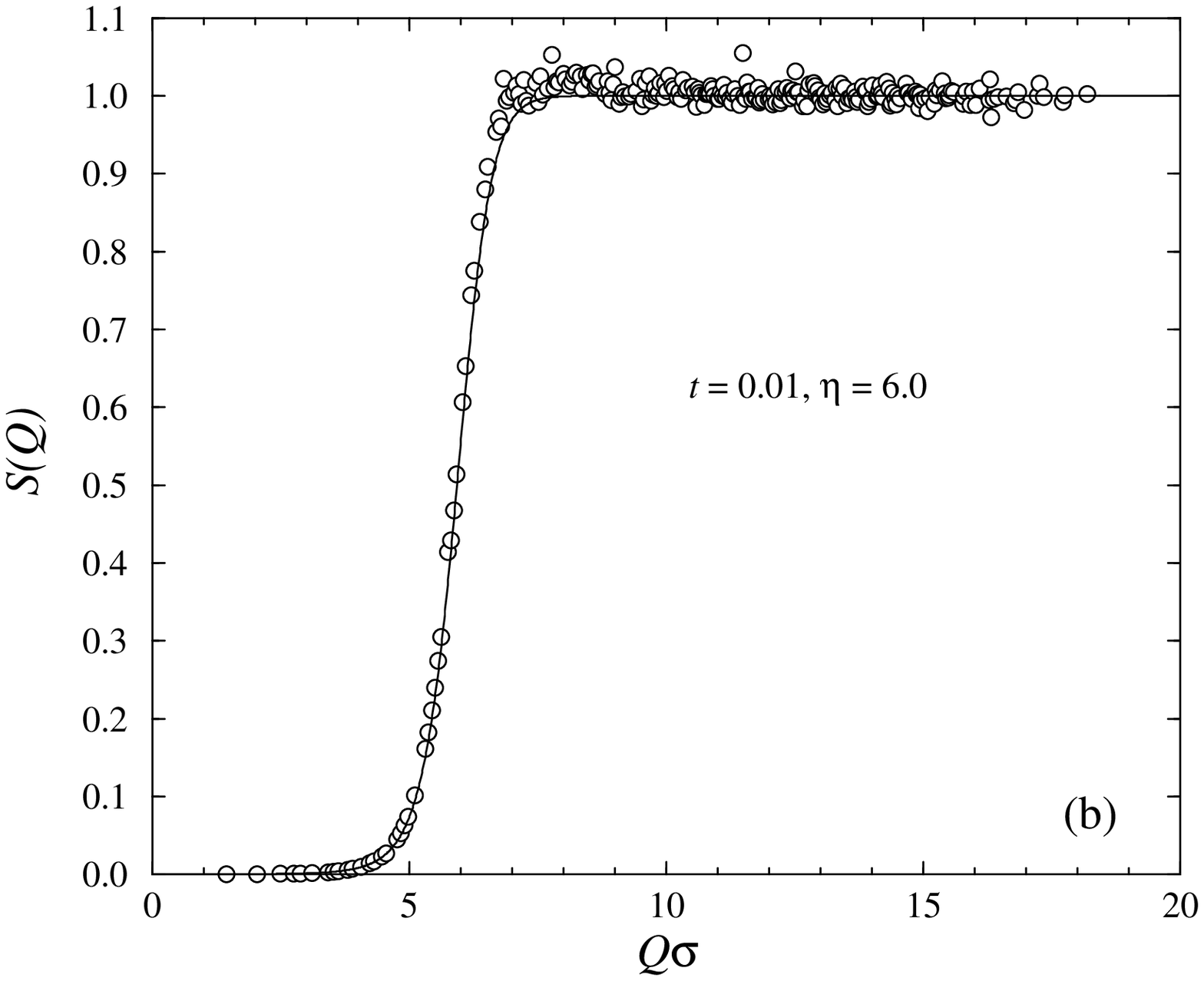}
   \end{minipage}
   \end{center}
   \caption{Same as Fig. \ref{xi0.6.plot} but for the Gaussian core model.}
\label{gcm.plot}
\end{figure}

These characteristic features for
the $Q^{+}$-class are not an artifact of the relatively high temperature 
chosen in the results of Fig.\ \ref{xi0.6.plot}. They persist 
even at extremely low temperatures, provided the density is high enough.
This has been amply demonstrated recently
for the case of the Gaussian core model, another member of the
$Q^{+}$-class \cite{lang:etal:gcm}. In order to stress this point
we present in Fig.\ \ref{gcm.plot} the $g(r)$ and $S(Q)$ of the
GCM at $t = 0.01$ and $\eta = 6.0$. Though $g(r)$ displays 
some structure up to $r \approx 2\sigma$, the structure factor
$S(Q)$ shows no signature of some kind of 
ordering \footnote{The small discrepancies between 
MFA and MC for $g(r)$ at small
$r$ values disappear as the density is increased. Perfect agreement
between theory and simulation can be achieved if, instead of the
MFA, one employs
the HNC or some more refined version of the MFA to calculate the
fluid structure, see Refs.\ \cite{lang:etal:gcm} and \cite{ard:mft:00}.}.
At any
arbitrarily small but finite temperature, a high enough density
can be found for which the MFA is valid and then the assumption
that a uniform phase exists leads consistently to a fluid which
has ideal-gas behavior, i.e., vanishingly small correlations.
These liquids are different from usual ideal gases in
that, e.g., their pressure $P$ and isothermal compressiblity $\chi_T$
scale respectively as $P \sim \rho^2$ and $\chi_T \sim t\rho^{-2}$.
Nevertheless, they are thermodynamically stable. Hence, 
for potentials in the $Q^{+}$-class, the equilibrium phase for
sufficiently high densities at arbitrarily small but finite
temperatures is the uniform fluid. 

\section{Generic phase diagrams}

We now turn to the opposite limit of the low temperature-low density part
of the phase diagram. There, following the original ideas of
Stillinger \cite{stillinger:76}, a HS mapping can be performed, as
follows. The Boltzmann factor $\exp[-\beta v(r)]$ of the potential
varies monotonically from the value $\exp(-\beta\varepsilon) \cong 0$
(since $\beta\varepsilon \gg 1$ there)
at $r = 0$ to unity at $r \rightarrow \infty$
and has a close resemblance to that of a hard sphere system. We can thus
define
an effective hard sphere diameter $\sigma_{\rm HS}$ through the relation:
\begin{equation}
\exp[-\beta v(\sigma_{\rm HS})] = 1/2.
\label{eff.hs}
\end{equation}
Writing $v(r) = \varepsilon\phi(r/\sigma)$ and using the fact
that $\phi(x)$ is a monotonic function in order to establish
that the inverse function $\phi^{-1}(x)$ exists,
we can rewrite eq.\ (\ref{eff.hs}) as:
\begin{equation}
\sigma_{\rm HS} = \sigma\phi^{-1}(t\ln 2).
\label{sigma.hs}
\end{equation}
We now use the known fact hard spheres freeze at
$\eta_{\rm HS} \cong 0.5$ together with 
eq.\ (\ref{sigma.hs}) above in oder to obtain the 
low temperature-low density freezing line of the system as:
\begin{equation}
t_{\rm f}(\eta) = \frac{1}{\ln 2}\phi\left(\left(2\eta\right)^{-1/3}\right).
\label{freeze.line}
\end{equation}
As the limit
$\phi(x) \to 0$ is attained for $x \to \infty$ only, it follows that
the low temperature-low density
freezing line of the systems goes to  
$\eta = 0$ at
$t_{\rm f} = 0$.
Eq.\ (\ref{freeze.line}) is valid for {\it all} potentials we
consider here; however, for $Q^{+}$-potentials, combining the
HS-like freezing at low temperatures and densities with the
fact that at high densities the fluid has to be stable, 
derived in the preceding section, we can
draw the conclusion that such systems must display reentrant melting
and an upper freezing temperature. 
\begin{figure}
\begin{center}
\includegraphics[width=6.0cm,height=8.0cm,angle=-90,clip]
{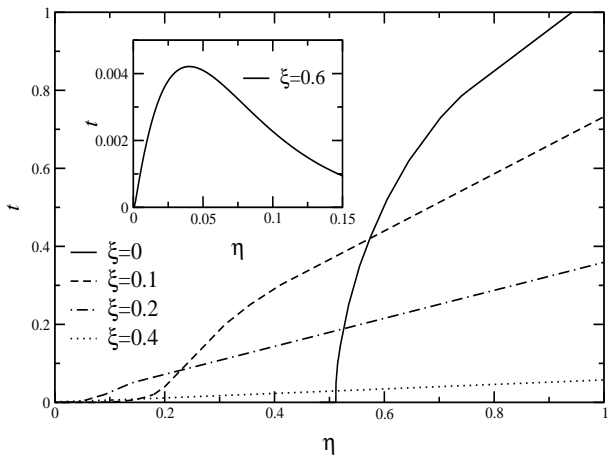}
\end{center}
\caption{The evolution of the phase diagram of $Q^{\pm}$-FDM's with $\xi$. To
         the right of the freezing lines the system
         is solid and to the left fluid.
         Inset: the phase diagram of a $Q^{+}$-FDM with $\xi > \xi_c$,
         obtained by solving
         the HNC and employing the Hansen-Verlet criterion.
         Below the bell-shaped
         curve the system is solid and above fluid.}
\label{phdg.plot}
\end{figure}

We have now taken eq.\ (\ref{freeze.line}) for
the low-$t$ and low-$\eta$ freezing line of the FDM and combined
it with the analytic expression at the opposite limit,
eq.\ (\ref{freezing.line}), in order to draw schematically the evolution
of the phase diagram of the FDM as a function of $\xi$, for
$\xi < \xi_c$. The results are shown in Fig.\ \ref{phdg.plot}.
With increasing $\xi$, the slopes of the high-$t$ freezing lines decrease;
at the limit $\xi \rightarrow 0$, corresponding to the PSM,
the low-$t$ freezing line approaches
the horizontal axis vertically, as is dictated by the fact that the
PSM becomes equivalent to the HS system there \cite{likos:pensph:98}.
In the inset of Fig.\ \ref{phdg.plot},
we show the phase diagram for a system with $\xi = 0.6 > \xi_c$,
showing reentrant melting behavior. The evolution of the
phase diagram from a clustering to a reentrant melting behavior can
be easily visualized from this picture.

Finally, it is important to point out
that Stillinger has proven that any system
interacting by means of a potential which (i) is differentiable at
least four times, (ii) vanishes strongly enough at infinity to 
be integrable and (iii) is +1 at the origin, will inevitably
lead to a reentrant melting phase diagram {\it under the assumption} that
the competing crystal structures have 
{\it single lattice site occupancy} \cite{stillinger:76}. We can
therefore now complete the statement and say that if a potential
belongs to the $Q^{+}$-class, then it will freeze into crystals
of single occupancy and then remelt upon increase of the density.
But if it belongs to the $Q^{\pm}$-class, then it will freeze
into a clustered solid at any temperature. Clustering appears therefore
to be the
crucial mechanism for crystal stabilization in these systems.
 
\section{Summary and concluding remarks}

To summarize, we have established a criterion for the topology 
of the phase diagrams resulting from repulsive, bounded interactions,
which is very simple in its formulation and states that if the Fourier
transform of the pair potential is positive-definite, then the system
shows reentrant melting but if not then it freezes at all temperatures,
into clustered crystals with multiply occupied sites. We have also
established that at temperatures exceeding the interaction strength
the mean-field theory is reliable at {\it all} densities and its accuracy
improves quickly with increasing temperature. 
We close with the remark that there is a
certain similarity between the ideas put forward here and
the considerations on freezing for systems featuring of {\it diverging}
interactions at {\it infinite} spatial
dimensions \cite{frisch:etal:prl:85,klein:frisch:jcp:86,frisch:percus:pra:87,bagchi:rice:jcp:88}. However, in the latter case, the direct correlation
function is given by the Mayer function $f(r) = \exp[-\beta v(r)] - 1$
of the interaction potential and not by $-\beta v(r)$ as in the case
at hand.
 
\acknowledgements

This work was supported by the \"Osterreichische Forschungsfond under Project
Nos.\ P11194-PHY and P13062-TPH.
A.\ L.\ acknowledges financial support
by the Deutsche Forschungsgemeinschaft within the
SFB 237.


\begin{thebibliography}{10}
\expandafter\ifx\csname bibnamefont\endcsname\relax
  \def\bibnamefont#1{#1}\fi
\expandafter\ifx\csname bibfnamefont\endcsname\relax
  \def\bibfnamefont#1{#1}\fi
\expandafter\ifx\csname url\endcsname\relax
  \def\url#1{\texttt{#1}}\fi
\expandafter\ifx\csname urlprefix\endcsname\relax\def\urlprefix{URL }\fi
\expandafter\ifx\csname bibinfo\endcsname\relax \def\bibinfo#1#2{#2}\fi
\expandafter\ifx\csname eprint\endcsname\relax \def\eprint#1{#1}\fi

\bibitem{mcconnell:etal:prl:93}
\bibinfo{author}{\bibfnamefont{G.~A.} \bibnamefont{McConnell}},
  \bibinfo{author}{\bibfnamefont{A.~P.} \bibnamefont{Gast}},
  \bibinfo{author}{\bibfnamefont{J.~S.} \bibnamefont{Huang}}, \bibnamefont{and}
  \bibinfo{author}{\bibfnamefont{S.~D.} \bibnamefont{Smith}},
  \bibinfo{journal}{Phys. Rev. Lett.} \textbf{\bibinfo{volume}{71}},
  \bibinfo{pages}{2102} (\bibinfo{year}{1993}).

\bibitem{watzlawek:etal:prl:99}
\bibinfo{author}{\bibfnamefont{M.}~\bibnamefont{Watzlawek}},
  \bibinfo{author}{\bibfnamefont{C.~N.} \bibnamefont{Likos}}, \bibnamefont{and}
  \bibinfo{author}{\bibfnamefont{H.}~\bibnamefont{L{\"o}wen}},
  \bibinfo{journal}{Phys. Rev. Lett.} \textbf{\bibinfo{volume}{82}},
  \bibinfo{pages}{5289} (\bibinfo{year}{1999}).

\bibitem{ard:peter:00}
\bibinfo{author}{\bibfnamefont{A.~A.} \bibnamefont{Louis}},
  \bibinfo{author}{\bibfnamefont{P.~G.} \bibnamefont{Bolhuis}},
  \bibinfo{author}{\bibfnamefont{J.-P.} \bibnamefont{Hansen}},
  \bibnamefont{and} \bibinfo{author}{\bibfnamefont{E.~J.} \bibnamefont{Meier}},
  \bibinfo{journal}{Phys. Rev. Lett.} \textbf{\bibinfo{volume}{85}},
  \bibinfo{pages}{2522} (\bibinfo{year}{2000}).

\bibitem{likos:ballauff}
\bibinfo{author}{\bibfnamefont{C.~N.} \bibnamefont{Likos}},
  \bibinfo{author}{\bibfnamefont{M.}~\bibnamefont{Schmidt}},
  \bibinfo{author}{\bibfnamefont{H.}~\bibnamefont{L{\"o}wen}},
  \bibinfo{author}{\bibfnamefont{M.}~\bibnamefont{Ballauff}},
  \bibinfo{author}{\bibfnamefont{D.}~\bibnamefont{P{\"o}tschke}},
  \bibnamefont{and} \bibinfo{author}{\bibfnamefont{P.}~\bibnamefont{Lindner}},
  \emph{\bibinfo{title}{{\it Soft interaction between dissolved dendrimers:
  theory and experiment}}}, \bibinfo{note}{submitted to Macromolecules (2000)}.

\bibitem{marquest:witten:89}
\bibinfo{author}{\bibfnamefont{C.}~\bibnamefont{Marquest}} \bibnamefont{and}
  \bibinfo{author}{\bibfnamefont{T.~A.} \bibnamefont{Witten}},
  \bibinfo{journal}{J. Phys. (Paris)} \textbf{\bibinfo{volume}{50}},
  \bibinfo{pages}{1267} (\bibinfo{year}{1989}).

\bibitem{likos:pensph:98}
\bibinfo{author}{\bibfnamefont{C.~N.} \bibnamefont{Likos}},
  \bibinfo{author}{\bibfnamefont{M.}~\bibnamefont{Watzlawek}},
  \bibnamefont{and}
  \bibinfo{author}{\bibfnamefont{H.}~\bibnamefont{L{\"o}wen}},
  \bibinfo{journal}{Phys. Rev. E} \textbf{\bibinfo{volume}{58}},
  \bibinfo{pages}{3135} (\bibinfo{year}{1998}).

\bibitem{stillinger:76}
\bibinfo{author}{\bibfnamefont{F.~H.} \bibnamefont{Stillinger}},
  \bibinfo{journal}{J. Chem. Phys.} \textbf{\bibinfo{volume}{65}},
  \bibinfo{pages}{3968} (\bibinfo{year}{1976}).

\bibitem{olaj:lantschbauer:77}
\bibinfo{author}{\bibfnamefont{O.~F.} \bibnamefont{Olaj}} \bibnamefont{and}
  \bibinfo{author}{\bibfnamefont{W.}~\bibnamefont{Lantschbauer}},
  \bibinfo{journal}{Ber. Bunsen-Ges. physik. Chem.}
  \textbf{\bibinfo{volume}{81}}, \bibinfo{pages}{985} (\bibinfo{year}{1977}).

\bibitem{grosberg:82}
\bibinfo{author}{\bibfnamefont{A.~Y.} \bibnamefont{Grosberg}},
  \bibinfo{author}{\bibfnamefont{P.~G.} \bibnamefont{Khalatur}},
  \bibnamefont{and} \bibinfo{author}{\bibfnamefont{A.~R.}
  \bibnamefont{Khokhlov}}, \bibinfo{journal}{Makromol. Chem. Rapid Commun.}
  \textbf{\bibinfo{volume}{3}}, \bibinfo{pages}{709} (\bibinfo{year}{1982}).

\bibitem{schaefer:baumgaertner:86}
\bibinfo{author}{\bibfnamefont{L.}~\bibnamefont{Sch{\"a}fer}} \bibnamefont{and}
  \bibinfo{author}{\bibfnamefont{A.}~\bibnamefont{Baumg{\"a}rtner}},
  \bibinfo{journal}{J. Phys. (Paris)} \textbf{\bibinfo{volume}{47}},
  \bibinfo{pages}{1431} (\bibinfo{year}{1986}).

\bibitem{krueger:etal:89}
\bibinfo{author}{\bibfnamefont{B.}~\bibnamefont{Kr{\"u}ger}},
  \bibinfo{author}{\bibfnamefont{L.}~\bibnamefont{Sch{\"a}fer}},
  \bibnamefont{and}
  \bibinfo{author}{\bibfnamefont{A.}~\bibnamefont{Baumg{\"a}rtner}},
  \bibinfo{journal}{J. Phys. (Paris)} \textbf{\bibinfo{volume}{50}},
  \bibinfo{pages}{3191} (\bibinfo{year}{1989}).

\bibitem{dautenhahn:hall:94}
\bibinfo{author}{\bibfnamefont{J.}~\bibnamefont{Dautenhahn}} \bibnamefont{and}
  \bibinfo{author}{\bibfnamefont{C.~K.} \bibnamefont{Hall}},
  \bibinfo{journal}{Macromolecules} \textbf{\bibinfo{volume}{27}},
  \bibinfo{pages}{5933} (\bibinfo{year}{1994}).

\bibitem{ard:finken:hansen:99}
\bibinfo{author}{\bibfnamefont{A.~A.} \bibnamefont{Louis}},
  \bibinfo{author}{\bibfnamefont{R.}~\bibnamefont{Finken}}, \bibnamefont{and}
  \bibinfo{author}{\bibfnamefont{J.-P.} \bibnamefont{Hansen}},
  \bibinfo{journal}{Europhys. Lett.} \textbf{\bibinfo{volume}{46}},
  \bibinfo{pages}{741} (\bibinfo{year}{1999}).

\bibitem{bolhuis:jcp:00}
\bibinfo{author}{\bibfnamefont{P.~G.} \bibnamefont{Bolhuis}},
  \bibinfo{author}{\bibfnamefont{A.~A.} \bibnamefont{Louis}},
  \bibinfo{author}{\bibfnamefont{J.-P.} \bibnamefont{Hansen}},
  \bibnamefont{and} \bibinfo{author}{\bibfnamefont{E.~J.} \bibnamefont{Meier}},
  \emph{\bibinfo{title}{Accurate effective pair potentials for polymer
  solutions}}, \bibinfo{note}{preprint, cond-mat/0009093, submitted to J. Chem.
  Phys. (2000)}.

\bibitem{fernaud:jcp:00}
\bibinfo{author}{\bibfnamefont{M.~J.} \bibnamefont{Fernaud}},
  \bibinfo{author}{\bibfnamefont{E.}~\bibnamefont{Lomba}}, \bibnamefont{and}
  \bibinfo{author}{\bibfnamefont{L.~L.} \bibnamefont{Lee}},
  \bibinfo{journal}{J. Chem. Phys.} \textbf{\bibinfo{volume}{112}},
  \bibinfo{pages}{810} (\bibinfo{year}{2000}).

\bibitem{schmidt:cecam:99}
\bibinfo{author}{\bibfnamefont{M.}~\bibnamefont{Schmidt}}, \bibinfo{journal}{J.
  Phys.: Condens. Matter} \textbf{\bibinfo{volume}{11}}, \bibinfo{pages}{10163}
  (\bibinfo{year}{1999}).

\bibitem{rosenfeld:etal:pensph}
\bibinfo{author}{\bibfnamefont{Y.}~\bibnamefont{Rosenfeld}},
  \bibinfo{author}{\bibfnamefont{M.}~\bibnamefont{Schmidt}},
  \bibinfo{author}{\bibfnamefont{M.}~\bibnamefont{Watzlawek}},
  \bibnamefont{and}
  \bibinfo{author}{\bibfnamefont{H.}~\bibnamefont{L{\"o}wen}},
  \bibinfo{journal}{Phys. Rev. E} \textbf{\bibinfo{volume}{62}},
  \bibinfo{pages}{5006} (\bibinfo{year}{2000}).

\bibitem{stillinger:weber:78}
\bibinfo{author}{\bibfnamefont{F.~H.} \bibnamefont{Stillinger}}
  \bibnamefont{and} \bibinfo{author}{\bibfnamefont{T.~A.} \bibnamefont{Weber}},
  \bibinfo{journal}{J. Chem. Phys.} \textbf{\bibinfo{volume}{68}},
  \bibinfo{pages}{3837} (\bibinfo{year}{1978}).

\bibitem{stillinger:weber:80}
\bibinfo{author}{\bibfnamefont{F.~H.} \bibnamefont{Stillinger}}
  \bibnamefont{and} \bibinfo{author}{\bibfnamefont{T.~A.} \bibnamefont{Weber}},
  \bibinfo{journal}{Phys. Rev. B} \textbf{\bibinfo{volume}{22}},
  \bibinfo{pages}{3790} (\bibinfo{year}{1980}).

\bibitem{stillinger:jcp:79}
\bibinfo{author}{\bibfnamefont{F.~H.} \bibnamefont{Stillinger}},
  \bibinfo{journal}{J. Chem. Phys.} \textbf{\bibinfo{volume}{70}},
  \bibinfo{pages}{4067} (\bibinfo{year}{1979}).

\bibitem{stillinger:prb:79}
\bibinfo{author}{\bibfnamefont{F.~H.} \bibnamefont{Stillinger}},
  \bibinfo{journal}{Phys. Rev. B} \textbf{\bibinfo{volume}{20}},
  \bibinfo{pages}{299} (\bibinfo{year}{1979}).

\bibitem{lang:etal:gcm}
\bibinfo{author}{\bibfnamefont{A.}~\bibnamefont{Lang}},
  \bibinfo{author}{\bibfnamefont{C.~N.} \bibnamefont{Likos}},
  \bibinfo{author}{\bibfnamefont{M.}~\bibnamefont{Watzlawek}},
  \bibnamefont{and}
  \bibinfo{author}{\bibfnamefont{H.}~\bibnamefont{L{\"o}wen}},
  \bibinfo{journal}{J. Phys.: Condens. Matter} \textbf{\bibinfo{volume}{24}},
  \bibinfo{pages}{5087} (\bibinfo{year}{2000}).

\bibitem{ruelle}
\bibinfo{author}{\bibfnamefont{D.}~\bibnamefont{Ruelle}},
  \emph{\bibinfo{title}{Statistical Mechanics}} (\bibinfo{publisher}{W. A.
  Benjamin}, \bibinfo{address}{New York}, \bibinfo{year}{1969}).

\bibitem{evans:79}
\bibinfo{author}{\bibfnamefont{R.}~\bibnamefont{Evans}}, \bibinfo{journal}{Adv.
  Phys.} \textbf{\bibinfo{volume}{28}}, \bibinfo{pages}{143}
  (\bibinfo{year}{1979}).

\bibitem{hansen:mcdonald}
\bibinfo{author}{\bibfnamefont{J.-P.} \bibnamefont{Hansen}} \bibnamefont{and}
  \bibinfo{author}{\bibfnamefont{I.~R.} \bibnamefont{McDonald}},
  \emph{\bibinfo{title}{Theory of Simple Liquids}}
  (\bibinfo{publisher}{Academic Press}, \bibinfo{address}{London},
  \bibinfo{year}{1986}), second ed.

\bibitem{ard:mft:00}
\bibinfo{author}{\bibfnamefont{A.~A.} \bibnamefont{Louis}},
  \bibinfo{author}{\bibfnamefont{P.~G.} \bibnamefont{Bolhuis}},
  \bibnamefont{and} \bibinfo{author}{\bibfnamefont{J.-P.}
  \bibnamefont{Hansen}}, \emph{\bibinfo{title}{{\it Mean field fluid behavior
  of the Gaussian core model}}}, \bibinfo{note}{preprint, cond-mat/0007062, to
  appear in Phys. Rev. E (2000)}.

\bibitem{grewe:klein:jmpa:77}
\bibinfo{author}{\bibfnamefont{N.}~\bibnamefont{Grewe}} \bibnamefont{and}
  \bibinfo{author}{\bibfnamefont{W.}~\bibnamefont{Klein}}, \bibinfo{journal}{J.
  Math. Phys.} \textbf{\bibinfo{volume}{64}}, \bibinfo{pages}{1729}
  (\bibinfo{year}{1977}).

\bibitem{grewe:klein:jmpb:77}
\bibinfo{author}{\bibfnamefont{N.}~\bibnamefont{Grewe}} \bibnamefont{and}
  \bibinfo{author}{\bibfnamefont{W.}~\bibnamefont{Klein}}, \bibinfo{journal}{J.
  Math. Phys.} \textbf{\bibinfo{volume}{64}}, \bibinfo{pages}{1735}
  (\bibinfo{year}{1977}).

\bibitem{klein:grewe:jcp:80}
\bibinfo{author}{\bibfnamefont{W.}~\bibnamefont{Klein}} \bibnamefont{and}
  \bibinfo{author}{\bibfnamefont{N.}~\bibnamefont{Grewe}}, \bibinfo{journal}{J.
  Chem. Phys.} \textbf{\bibinfo{volume}{72}}, \bibinfo{pages}{5456}
  (\bibinfo{year}{1980}).

\bibitem{klein:brown:jcp:81}
\bibinfo{author}{\bibfnamefont{W.}~\bibnamefont{Klein}} \bibnamefont{and}
  \bibinfo{author}{\bibfnamefont{A.~C.} \bibnamefont{Brown}},
  \bibinfo{journal}{J. Chem. Phys.} \textbf{\bibinfo{volume}{74}},
  \bibinfo{pages}{6960} (\bibinfo{year}{1981}).

\bibitem{klein:etal:physica:94}
\bibinfo{author}{\bibfnamefont{W.}~\bibnamefont{Klein}},
  \bibinfo{author}{\bibfnamefont{H.}~\bibnamefont{Gould}},
  \bibinfo{author}{\bibfnamefont{R.~A.} \bibnamefont{Ramos}},
  \bibinfo{author}{\bibfnamefont{I.}~\bibnamefont{Clejan}}, \bibnamefont{and}
  \bibinfo{author}{\bibfnamefont{A.~I.} \bibnamefont{Mel'cuk}},
  \bibinfo{journal}{Physica A} \textbf{\bibinfo{volume}{205}},
  \bibinfo{pages}{738} (\bibinfo{year}{1994}).

\bibitem{stillinger:stillinger:97}
\bibinfo{author}{\bibfnamefont{F.~H.} \bibnamefont{Stillinger}}
  \bibnamefont{and} \bibinfo{author}{\bibfnamefont{D.~K.}
  \bibnamefont{Stillinger}}, \bibinfo{journal}{Physica A}
  \textbf{\bibinfo{volume}{244}}, \bibinfo{pages}{358} (\bibinfo{year}{1997}).

\bibitem{hansen:verlet:69}
\bibinfo{author}{\bibfnamefont{J.-P.} \bibnamefont{Hansen}} \bibnamefont{and}
  \bibinfo{author}{\bibfnamefont{L.}~\bibnamefont{Verlet}},
  \bibinfo{journal}{Phys. Rev.} \textbf{\bibinfo{volume}{184}},
  \bibinfo{pages}{151} (\bibinfo{year}{1969}).

\bibitem{hansen:schiff:73}
\bibinfo{author}{\bibfnamefont{J.-P.} \bibnamefont{Hansen}} \bibnamefont{and}
  \bibinfo{author}{\bibfnamefont{D.}~\bibnamefont{Schiff}},
  \bibinfo{journal}{Mol. Phys.} \textbf{\bibinfo{volume}{25}},
  \bibinfo{pages}{1281} (\bibinfo{year}{1973}).

\bibitem{watzlawek:etal:jpcm:98}
\bibinfo{author}{\bibfnamefont{M.}~\bibnamefont{Watzlawek}},
  \bibinfo{author}{\bibfnamefont{H.}~\bibnamefont{L{\"o}wen}},
  \bibnamefont{and} \bibinfo{author}{\bibfnamefont{C.~N.} \bibnamefont{Likos}},
  \bibinfo{journal}{J. Phys.: Condens. Matter} \textbf{\bibinfo{volume}{10}},
  \bibinfo{pages}{8189} (\bibinfo{year}{1998}).

\bibitem{frisch:etal:prl:85}
\bibinfo{author}{\bibfnamefont{H.~L.} \bibnamefont{Frisch}},
  \bibinfo{author}{\bibfnamefont{N.}~\bibnamefont{Rivier}}, \bibnamefont{and}
  \bibinfo{author}{\bibfnamefont{D.}~\bibnamefont{Wyler}},
  \bibinfo{journal}{Phys. Rev. Lett.} \textbf{\bibinfo{volume}{54}},
  \bibinfo{pages}{2061} (\bibinfo{year}{1985}).

\bibitem{klein:frisch:jcp:86}
\bibinfo{author}{\bibfnamefont{W.}~\bibnamefont{Klein}} \bibnamefont{and}
  \bibinfo{author}{\bibfnamefont{H.~L.} \bibnamefont{Frisch}},
  \bibinfo{journal}{J. Chem. Phys.} \textbf{\bibinfo{volume}{84}},
  \bibinfo{pages}{968} (\bibinfo{year}{1986}).

\bibitem{frisch:percus:pra:87}
\bibinfo{author}{\bibfnamefont{H.~L.} \bibnamefont{Frisch}} \bibnamefont{and}
  \bibinfo{author}{\bibfnamefont{J.~K.} \bibnamefont{Percus}},
  \bibinfo{journal}{Phys. Rev. A} \textbf{\bibinfo{volume}{35}},
  \bibinfo{pages}{4696} (\bibinfo{year}{1987}).

\bibitem{bagchi:rice:jcp:88}
\bibinfo{author}{\bibfnamefont{B.}~\bibnamefont{Bagchi}} \bibnamefont{and}
  \bibinfo{author}{\bibfnamefont{S.~A.} \bibnamefont{Rice}},
  \bibinfo{journal}{J. Chem. Phys.} \textbf{\bibinfo{volume}{88}},
  \bibinfo{pages}{1177} (\bibinfo{year}{1988}).

\end{thebibliography}
\end{document}